\documentclass[12pt,a4paper]{article}

\usepackage{amsmath,amssymb,amsfonts}
\usepackage{geometry}
\usepackage{graphicx}          % was missing in v1 although \includegraphics was used
\usepackage{hyperref}
\usepackage{enumitem}
\usepackage{bm}
\usepackage{xcolor}
\usepackage{tikz}
\usepackage{multirow}
\usepackage{authblk}
\usepackage{booktabs}
\usepackage{caption}
\usetikzlibrary{arrows.meta,positioning,shapes.geometric,decorations.pathmorphing,calc,fit,backgrounds}

\newcommand{\Mpl}{M_{\rm Pl}}
\newcommand{\KK}{\mathcal{K}}
\newcommand{\UU}{\mathcal{U}}
\newcommand{\NN}{\mathcal{N}}
\newcommand{\dln}{\mathrm{d}\ln a}
\newcommand{\dd}{\mathrm{d}}
\newcommand{\aM}{\alpha_{\!M}}
\newcommand{\lF}{\lambda_F}
\newcommand{\lU}{\lambda_U}
\newcommand{\GF}{\Gamma_{\!F}}
\newcommand{\GU}{\Gamma_{\!U}}
\newcommand{\Om}{\Omega_{\rm m}}

\title{Cosmographic Reconstruction of the Quintessence Potential\\[4pt]
in Scalar-Tensor Gravity}
\author[ ]{G. Antonopoulos\thanks{\texttt{gio.antonopoulos12@gmail.com}}}
\author[ ]{L. Perivolaropoulos\thanks{\texttt{leandros@uoi.gr}}}
\affil{Department of Physics, University of Ioannina, Greece}
\date{\today}

\begin{document}
\maketitle

\begin{abstract}
\noindent
We generalise the cosmographic reconstruction programme of
Chakraborty, Dunsby \& Scherrer~\cite{CDS2026} to scalar-tensor
gravity. Working in the Jordan frame with a general non-minimal
coupling $F(\Phi)$, we derive exact closed-form expressions for the
potential slope $\lambda_U$, the coupling slope $\lambda_F$, and their
curvature parameters $\Gamma_U$ and $\Gamma_F$ in terms of the
cosmographic parameters $(q,j,s)$, the Planck-mass running
$\alpha_M = d\ln F/d\ln a$ and its derivatives.  We express $\alpha_M$ and its derivatives in terms of  a set of coefficients
$g_n$ describing the time variation of the gravitational constant
$G\propto F^{-1}$, casting all four quantities in fully observable form. We then
evaluate the reconstruction against current data, including DESI~DR2
baryon acoustic oscillation measurements and Lunar Laser Ranging
constraints on $\dot G/G$, propagating the cosmographic uncertainties
through to the reconstructed potential. All expressions reduce
analytically to those of Ref.~\cite{CDS2026} in the minimally-coupled
limit, which we verify symbolically. We find that the potential
slope $\lambda_U$ is recovered at the few-tenths level while the
curvature $\Gamma_U$ is essentially unconstrained. We further show
that the positivity of the scalar kinetic term is equivalent to
$w_{\rm eff,0}\ge-1$, so that the region of cosmographic parameter
space closed to minimally-coupled quintessence is exactly the phantom
region, and that the non-minimal coupling reopens it through a single
coefficient $g_2$ of the gravitational tower. Because Lunar Laser
Ranging forces $|g_1|\lesssim2\times10^{-4}$ while the reconstruction
is sensitive to $g_2\sim O(0.1)$, the framework presupposes that
$\alpha_M$ is passing through zero at the present epoch, as expected
on the Damour--Nordtvedt least-coupling attractor.
\end{abstract}

%=====================================================
\section{Introduction}\label{sec:intro}
%=====================================================
The discovery of the accelerating
expansion of the
Universe~\cite{Riess1998,Perlmutter1999}
established the nature of dark energy as one of the
central problems of modern cosmology.
The standard $\Lambda$CDM model, in which
dark energy is a cosmological constant
$\Lambda$ with equation of state
$w = -1$, provides an excellent fit to a
broad range of observations, from the
cosmic microwave background~\cite{Planck2018}
to baryon acoustic oscillations and
Type~Ia supernovae.  Despite this  success, the assumption of a pure cosmological constant remains deeply problematic.
Theoretically, its observed value is many orders of magnitude smaller than naive vacuum-energy estimates~\cite{Weinberg1989}, requiring severe fine-tuning to explain its comparability to the present-day matter density.
Independent of these theoretical concerns,
recent
baryon acoustic oscillation data from
DESI~\cite{DESI2024VI,DESIDR2}, combined with
supernova compilations, suggest  a preference for
 dynamical dark energy
over a pure cosmological constant.

The simplest dynamical alternative  is
quintessence, a minimally coupled scalar
field $\Phi$ slowly rolling in a
potential
$U(\Phi)$~\cite{CaldwellDaveSteinhardt1998,
Tsujikawa2013}. A further extension
introduces a non-minimal coupling between the scalar field and gravity. In these theories the scalar field plays
a dual role: it can simultaneously drive
the late-time acceleration and mediate a
time variation of the gravitational
coupling. Such scalar-tensor theories
have a long history, beginning with the
 Brans--Dicke theory~\cite{BransDicke1961}, and
encompass $f(R)$ gravity in its
scalar-field formulation as well as the
broad Horndeski
class~\cite{Horndeski1974,Kobayashi2019}, the most general scalar-tensor theory with second-order field equations.
 Much of this parameter space is now sharply constrained: the near-simultaneous arrival of GW170817~\cite{Abbott2017detection} and its gamma-ray counterpart GRB~170817A~\cite{Goldstein2017,Savchenko2017} fixed the gravitational-wave speed to that of light, restricting the broader Horndeski class~\cite{Creminelli2017,Ezquiaga2017,Baker2017}. The minimal scalar-tensor sector with a canonical kinetic term, which we adopt here, remains viable.

A central goal of dark-energy phenomenology
is to reconstruct the dynamics of the
scalar sector directly from data, with
minimal model assumptions. Common approaches parametrise the equation of state $w(z)$~\cite{ChevallierPolarski2001,Linder2003} and infer the potential indirectly~\cite{HutererTurner1999}. A more direct route exploits cosmography, the model-independent expansion of the late-time scale factor~\cite{Visser2004,CattoenVisser2007}, which has seen renewed interest following the DESI results~\cite{Luongo:2024,Pourojaghi:2025,Mishra:2026}. Building on
this approach, Chakraborty, Dunsby and
Scherrer~\cite{CDS2026} recently showed
that for  quintessence
the potential slope $\lambda \equiv -(dU/d\Phi)/U$
 and curvature
$\Gamma \equiv (d^2U/d\Phi^2)\,U/(dU/d\Phi)^2$ can be expressed
exactly in terms of the cosmographic
parameters and the dark-energy density
fraction, enabling a direct Taylor
reconstruction of $U(\Phi)$ around the
present-day field value without reference
to $w$; a companion analysis specialises the
construction to nearly flat thawing
potentials~\cite{CDS2026b}.

In this work we extend  the Chakraborty--Dunsby--Scherrer reconstruction  to scalar-tensor gravity. The non-minimal
coupling $F(\Phi)$ is described by the Planck-mass running
$\alpha_M = d\ln F/d\ln a$ of Bellini and Sawicki~\cite{Bellini2014}.
We introduce a gravitational tower
$g_n \equiv d^n \ln G/d(\ln a)^n$, the logarithmic derivatives of the
coupling $G \propto F^{-1}$, which map $\alpha_M$ and its higher
derivatives onto the observable $\dot G/G$ and its successive time
derivatives. Combining this gravitational tower with the cosmographic parameters
$(q, j, s)$, we derive exact algebraic expressions for the coupling
slope $\lambda_F$, the potential slope $\lambda_U$, and their curvature
parameters $\Gamma_F$ and $\Gamma_U$. We then evaluate these expressions against recent data, including DESI~DR2 baryon acoustic oscillation measurements and Lunar Laser Ranging constraints on $\dot{G}/G$, to reconstruct the potential $U(\Phi)$ and the coupling $F(\Phi)$ near the present-day field value $\Phi_0$. Throughout we propagate the
quoted cosmographic uncertainties, which turn out to control the
usefulness of the reconstruction far more than the gravitational sector does.

The paper is organised as follows.
Section~\ref{sec:framework} sets up the
scalar-tensor action, introduces the cosmographic parameters $(q,j,s)$ and the identities relating them, and derives the dimensionless constraint equations.
Section~\ref{sec:Derivation} derives exact
expressions for $\lU$, $\lF$, $\GU$ and
$\GF$ in terms of cosmographic parameters
and the Planck-mass running $\aM$ together with its
derivatives, and
rewrites them in terms of the directly
observable gravitational tower $g_n$.
Section~\ref{sec:reconstruction} presents
the Taylor reconstruction of $U(\Phi)$
and $F(\Phi)$ and evaluates the results
against current cosmological data.
Section~\ref{sec:summary} summarises
the results. Appendix~\ref{app:cosmography_compilation} compiles recent
cosmographic constraints and
Appendix~\ref{app:gtower} gives the recursion generating the
gravitational tower.

%=====================================================
\section{Framework}\label{sec:framework}
%=====================================================
In this section we set up the scalar-tensor framework underlying the reconstruction. We first present the action and background field equations, then introduce the cosmographic parameters and the identities relating them, and finally cast the background system in the dimensionless form used throughout the paper.
\subsection{Action and field equations}

We work in the Jordan frame with the
action~\cite{BransDicke1961,Amendola2010}
\begin{equation}\label{eq:action}
  S = \int \dd^4 x\,\sqrt{-g}\left[
    \frac{1}{2}\,F(\Phi)\,R
    - \frac{1}{2}\,\partial_{\mu}\Phi\partial^{\mu}\Phi
    - U(\Phi)\right] + S_{\rm m}\,,
\end{equation}
where we have set $8\pi G_N = 1$ and the
kinetic term is canonically normalised. In these units $\Phi$ is
measured in reduced Planck masses, $\Mpl \equiv (8\pi G_N)^{-1/2} = 1$,
a convention we use for all field displacements below. The
matter sector $S_{\rm m}$ is minimally
coupled to the metric $g_{\mu\nu}$ and
contains no direct interaction with $\Phi$,
so the weak equivalence principle is
 satisfied. We take the matter sector to be pressureless, $\rho_m \propto a^{-3}$,
neglecting radiation at the late times of interest. Theoretical stability
further requires $F(\Phi) > 0$ to avoid
ghost instabilities in the graviton
sector~\cite{EspositoFarese2001}. Throughout,
 an overdot denotes
differentiation with respect to cosmic time
$t$, while a prime denotes differentiation
with respect to $\ln a$.

The non-minimal coupling $F(\Phi)$ implies
a time-varying effective gravitational
constant $G(a) \equiv 1/F(a)$. The  quantity probed   by local
tests such as Lunar Laser Ranging~\cite{Hofmann2018, Biskupek2021}, however, differs from this; the scalar field mediates an
additional force, so that~\cite{Boisseau2000}

\begin{equation}\label{eq:Geff}
  G_{\rm eff}(t) = \frac{1}{F(t)}\,
  \frac{F(t) + 2\,(\dd F/\dd\Phi)^2}
       {F(t) + \tfrac{3}{2}\,
        (\dd F/\dd\Phi)^2}
  \;\simeq\;G(t)\,,
\end{equation}
where  the second equality holds at low redshifts in the limit $(\dd F/\dd\Phi)^2 \ll F$, in which the
two coincide, $G_{\rm eff} \simeq G$. We verify in
Sec.~\ref{sec:ref_values} that this limit is very well satisfied
for the parameter values considered here, so that the LLR bound may
consistently be read as a bound on $\dot G/G$.

For a spatially flat
Friedmann--Lema\^itre--Robertson--Walker
background with homogeneous $\Phi$,
variation of the action yields the modified
Friedmann equation
\begin{equation}\label{eq:friedmann}
  3FH^2 = \rho_{\rm m}
    + \tfrac{1}{2}\dot\Phi^2
    + U - 3H\dot F\,,
\end{equation}
together with the Raychaudhuri equation
obtained by combining the $(0,0)$ and
$(i,j)$ field equations,
\begin{equation}\label{eq:raych}
  -2F\dot H = \rho_{\rm m} + \dot\Phi^2
    + \ddot F - H\dot F\,.
\end{equation}
The scalar field itself obeys the
Klein--Gordon equation
\begin{equation}\label{eq:KG}
  \ddot\Phi + 3H\dot\Phi
    + \frac{\dd U}{\dd\Phi}
    = \frac{1}{2}\,
      \frac{\dd F}{\dd\Phi}\,R\,,
\end{equation}
where the Ricci scalar is
$R = 6(1-q)H^2$ with $q$ the deceleration
parameter.
The reconstruction programme of
Sec.~\ref{sec:Derivation} proceeds
entirely from Eqs.~\eqref{eq:friedmann}
and~\eqref{eq:raych}, without invoking
the Klein--Gordon equation. This entails no loss of information:
given the contracted Bianchi identity and the separate conservation
of the pressureless matter sector, $\dot\rho_{\rm m} + 3H\rho_{\rm m}
= 0$, Eq.~\eqref{eq:KG} follows from Eqs.~\eqref{eq:friedmann}
and~\eqref{eq:raych} and is therefore redundant for the background
reconstruction.

%-----------------------------------------------------

\subsection{Cosmographic parameters and identities}
\label{sec:cosmography}
%------------------------------------------------------
Cosmography characterises the late-time expansion history through successive
derivatives of the scale factor~\cite{Visser2004,Sahni2003}. Normalising the
scale factor to the present epoch, $a(t_0) = 1$, and expanding $a(t)$ about
$t_0$ as a Taylor series,
\begin{equation}
  a(t) \simeq 1 + \dot a(t_0)\,(t-t_0)
  + \frac{\ddot a(t_0)}{2}\,(t-t_0)^2
  + \frac{\dddot a(t_0)}{6}\,(t-t_0)^3
  + \frac{\ddddot a(t_0)}{24}\,(t-t_0)^4
  + \dots,
  \label{eq:taylor-raw}
\end{equation}
one defines the cosmographic parameters $(H,q,j,s)$,
\begin{equation}
  H \equiv \frac{\dot a}{a}, \qquad
  q \equiv -\frac{\ddot a}{aH^2}, \qquad
  j \equiv \frac{\dddot a}{aH^3}, \qquad
  s \equiv \frac{\ddddot a}{aH^4},
  \label{eq:cosmographic-defs}
\end{equation}
known in the literature as the Hubble, deceleration, jerk and snap parameters.
In terms of these, the expansion~\eqref{eq:taylor-raw} takes the standard form
\begin{equation}
  a(t) \simeq 1 + H_0\,(t-t_0)
  - \frac{1}{2}q_0 H_0^2\,(t-t_0)^2
  + \frac{1}{3!}j_0 H_0^3\,(t-t_0)^3
  + \frac{1}{4!}s_0 H_0^4\,(t-t_0)^4
  + \dots.
  \label{eq:scale-factor-expansion}
\end{equation}
The Hubble parameter $H$ indicates whether the Universe expands or contracts.
In an expanding Universe the sign of $q$ distinguishes accelerated from
decelerated expansion, and the transition between the two epochs is the point
at which $q$  changes sign. The jerk $j$ governs how $q$
evolves through this point: since the present expansion is accelerating
($q_0 < 0$), a positive $j_0$ is consistent with a deceleration-to-acceleration
transition in the past~\cite{Pourojaghi2021}. In the standard $\Lambda$CDM
model $j_0 = 1$, so a measured departure of the jerk from unity would indicate
dynamics beyond a cosmological constant. The converse does not hold: as
emphasised in Refs.~\cite{CDS2026,Chakraborty2025}, $j=1$ alone does not
imply a flat potential, since $\lambda_U$ vanishes only if in addition
$\Omega_{\rm m} = \tfrac{2}{3}(1+q)$; a background with $j=1$ and
$w = -a^3/(a^3+C)$ is degenerate with $\Lambda$CDM cosmographically while
differing from it at the perturbative level.

The cosmographic parameters are  connected through the following identities
\begin{equation}
  q' = q(1+2q) - j, \qquad j' = j(2+3q) + s,
  \label{eq:cosmographic-identities}
\end{equation}
which, together with the matter evolution $\Omega_m' = (2q-1)\Omega_m$, are used
repeatedly in Sec.~\ref{sec:Derivation}. The set $(q,j,s)$ is the
cosmographic tower which, together with the gravitational tower introduced in
Sec.~\ref{sec:obs}, drives the reconstruction that follows.

The cosmographic series can in principle be extended to higher orders by
defining further parameters, but the reconstruction developed below requires
the expansion only up to the snap; we therefore restrict to $(q,j,s)$.
Extending further would in any case be of limited observational value, since
even the snap parameter $s$ is only weakly constrained by current data. A
further practical limitation is that, re-expressed in terms of the redshift
$1+z = 1/a$, the series converges only for $|z| \lesssim 1$, so its reliability
degrades when applied at higher redshifts, beyond the local Universe. This has
motivated reparametrisations designed to improve convergence over the full
expansion history, such as the $y = z/(1+z)$
variable~\cite{CattoenVisser2007}, together with Pad\'e and Chebyshev
polynomials~\cite{Aviles2014,Capozziello2018}. We use the cosmographic
parameters as inputs from such analyses in Sec.~\ref{sec:reconstruction}.

\subsection{Background dynamics in dimensionless form}

We introduce the dimensionless fractional
energy densities
\begin{equation}\label{eq:dimless-energy}
  \KK \equiv \frac{\dot\Phi^2}{6H^2}\,, \qquad
  \UU \equiv \frac{U}{3H^2}\,, \qquad
  \Om \equiv \frac{\rho_{\rm m}}{3H^2}\,,
\end{equation}
together with the dimensionless field
velocity $v \equiv \dot\Phi/H$, so that
$\KK = v^2/6$. The dynamics of the coupling are carried by its slope
$\lF = (\dd F/\dd\Phi)/F$, the scalar-tensor analogue of the potential
slope $\lambda_U$. Through the chain rule it is connected to the Planck-mass
running of Bellini and Sawicki~\cite{Bellini2014},
\begin{equation}\label{eq:alphaM}
  \aM \equiv \frac{\dot F}{HF} = \frac{\dd\ln F}{\dln} = \lambda_F\,v\,,
\end{equation}
which tracks the time variation of the gravitational constant $G = 1/F$
introduced in Eq.~\eqref{eq:Geff}.

We normalise $F = 1$, absorbing a constant into the field redefinition.
This fixes the present value of $F$ but not its derivatives: since
$\dd F/\dln = F\aM$, the quantities $\aM$, $\aM{'}$, and higher  derivatives remain
nonzero and carry the dynamical content of the sector. We therefore keep $F$
explicit in any relation that is differentiated further, imposing $F=1$ only
afterwards. This bookkeeping matters: in Eqs.~\eqref{eq:U_background}
and~\eqref{eq:v2_background} below the factor $F$ multiplies the scalar
contribution but not the matter term, so $F' = F\aM$ generates
$O(\aM)$ contributions that are lost if $F=1$ is imposed before
differentiating. One such term is retained explicitly in
Eq.~\eqref{eq:Nprime}.

Dividing the Friedmann
equation~\eqref{eq:friedmann} by $3H^2$
yields the Friedmann constraint
\begin{equation}\label{eq:Fr_constraint}
  \KK + \UU = F\,(1 + \aM) - \Om\,.
\end{equation}
The Raychaudhuri
equation~\eqref{eq:raych} requires
$\ddot F$, which follows from
differentiating Eq.~\eqref{eq:alphaM}
\begin{equation}\label{eq:Fddot}
  \frac{\ddot F}{F}
    = H^2\!\left[\aM{'} - \aM(1+q)\,
      + \aM^2\right].
\end{equation}
Substituting into Eq.~\eqref{eq:raych},
dividing by $3H^2$, and eliminating $\Om$
via Eq.~\eqref{eq:Fr_constraint} gives
the Raychaudhuri constraint
\begin{equation}\label{eq:Ray_constraint}
  \frac{3(\KK - \UU)}{F} = (2+\aM)\,q
    - 1 - \aM - \aM^2 - \aM{'}\,.
\end{equation}

Solving for $\UU$ and $v^2 = 6\KK$ yields

\begin{align}
  6\,\UU &= F\big[(2 + \aM)(2 + \aM - q) + \aM{'}\big] - 3\,\Om\,,
  \label{eq:U_background}\\[2pt]
  v^2 &= F\big[2(1 + q) + \aM(2 + q - \aM) - \aM{'}\big] - 3\,\Om\,.
  \label{eq:v2_background}
\end{align}
These expressions, together with the cosmographic identities~\eqref{eq:cosmographic-identities} and the derivatives of $\aM$
 up to third order in $\ln a$,
enter the reconstruction of Sec.~\ref{sec:Derivation}.

Equation~\eqref{eq:v2_background} already carries a strong consistency
requirement. Reality of the scalar field demands $v^2 > 0$, which in the
minimally-coupled limit reduces to the purely cosmographic bound
\begin{equation}\label{eq:Omcrit}
  \Om \;<\; \tfrac{2}{3}\,(1+q_0)\,,
\end{equation}
independent of $j$, $s$ and of the potential. This inequality has a direct
physical reading. Writing the background in terms of an effective
dark-energy equation of state through $q = \tfrac12 + \tfrac32 w\,\Omega_\phi$
\cite{CDS2026}, Eq.~\eqref{eq:v2_background} at $\aM=0$ becomes exactly
\begin{equation}\label{eq:v2_w}
  v^2 = 3\,\Omega_\phi\,(1 + w)\,,
\end{equation}
so $v^2>0$ is precisely the statement $w_{\rm eff,0} \ge -1$, and
Eq.~\eqref{eq:Omcrit} is the condition that the background not be phantom.
The region closed to minimally-coupled quintessence is therefore not an
artefact of the reconstruction but the phantom region itself. Restoring the
coupling, $v^2 = 2(1+q) - 3\Om + g_2$ for $|g_1|\ll1$, so a positive $g_2$
admits $w_{\rm eff,0} < -1$ with a healthy, non-ghost scalar --- the
well-known ability of scalar-tensor dark energy to cross the phantom
divide~\cite{Boisseau2000,EspositoFarese2001}. This is quantified in
Sec.~\ref{sec:results}.

%=====================================================
\section{Reconstruction of \texorpdfstring{$\lU$, $\lF$, $\GU$, and $\GF$}{lambda\_U, lambda\_F, Gamma\_U, and Gamma\_F}}\label{sec:Derivation}
In this section we derive exact algebraic expressions for the four
 slope and curvature parameters $\lU$, $\lF$, $\GU$ and $\GF$ in terms of the
cosmographic parameters $(q, j, s)$, the matter density parameter $\Om$ and the Planck-mass running
$\aM$  together with its derivatives up to
third order in $\ln a$. The derivation proceeds in two
stages: the potential slope $\lU$ is  obtained by differentiating the Raychaudhuri constraint
\eqref{eq:Ray_constraint}, while the curvature parameters $\GU$
and $\GF$ require one further differentiation. In
Sec.~\ref{sec:obs} the theoretical $\aM$-tower is then mapped onto
the observable $g_n$ tower, giving fully observable forms.

%-----------------------------------------------------
\subsection{Potential slope \texorpdfstring{$\lambda_U$}{lambda\_U}}\label{sec:routeII}
%-----------------------------------------------------
The first quantity we reconstruct is the
potential slope $\lambda_U = -(\dd U/\dd\Phi)/U$. We
derive it by differentiating the
Raychaudhuri
constraint~\eqref{eq:Ray_constraint} with
respect to $\ln a$ and using the
cosmographic identity
$q{'} = q(1+2q) - j$. This yields
\begin{equation}\label{eq:dRay}
  3(\KK{'} - \UU') - 3\aM(\KK - \UU)\,
  = (2 + \aM)\,q'
  - \aM'\,(1 + 2\aM - q) - \aM{''}\,,
\end{equation}
where the $\aM$ term on the left reflects the $F$-dependence retained in
Eq.~\eqref{eq:Ray_constraint}, and $\KK-\UU$ is supplied by that same
constraint.
The derivative $\UU'$ follows from the definition of $\UU$
in Eq.~\eqref{eq:dimless-energy}
\begin{equation}\label{eq:Uprime}
  \UU' = \UU\!\left[2(1+q) - \lU\,v\right],
\end{equation}
which introduces $\lambda_U$. The derivative $\KK'$ is obtained by
differentiating the Friedmann
constraint~\eqref{eq:Fr_constraint} and
using $\Om' = (2q-1)\,\Om$, giving
\begin{equation}\label{eq:Kprime}
  \KK' = \aM'+\aM+\aM^{2} - (2q-1)\,\Om - \UU'\,,
\end{equation}
so that
\begin{equation}\label{eq:KminusUprime}
  \KK' - \UU' = \aM+\aM^{2}+\aM' - (2q-1)\,\Om
    - 2\,\UU'\,.
\end{equation}
Substituting Eqs.~\eqref{eq:Uprime} and~\eqref{eq:KminusUprime} into
Eq.~\eqref{eq:dRay}, and using Eq.~\eqref{eq:Ray_constraint} at $F=1$ for
$\KK-\UU$, we obtain the closed-form expression
\begin{equation}\label{lambda_u}
  \lU = \frac{\NN}{6\,v\,\UU}\,,
\end{equation}
where $\UU$  and $v$ are given by
Eqs.~\eqref{eq:U_background},\eqref{eq:v2_background} and
\begin{equation}\label{eq:calN}
  \NN \equiv (2+\aM)\!\Big[(2+\aM)(2-\aM)+3(1+\aM)\,q - j\Big]
    +(3q-2-3\aM)\,\aM' - 9\,\Om - \aM{''}\,.
\end{equation}
Equation~\eqref{lambda_u}  expresses
$\lambda_U$ as an exact algebraic
function of the  parameters
$(q,j,\Om)$ along with the Planck-mass running
 and its first two derivatives
$(\aM, \aM', \aM{''})$. In the
minimally-coupled limit
$\aM = \aM' = \aM{''} = 0$ this expression reduces to
\begin{equation}\label{eq:lU_mincoupled}
  \lU \;\xrightarrow{\;\aM\to0\;}\;
  \frac{2\left[4 + 3q - j\right] - 9\,\Om}
       {\sqrt{2(1+q)-3\Om}\;\left[2(2-q)-3\Om\right]}\,,
\end{equation}
which, after substituting $\Omega_m = 1 - \Omega_\phi$, is exactly Eq.~(11)
of Ref.~\cite{CDS2026}. The dictionary between the two sets of variables is
\begin{equation}\label{eq:dictionary}
  \NN \;\leftrightarrow\; 9\Omega_\phi + 6q - 2j - 1\,,\qquad
  6\,\UU \;\leftrightarrow\; 3\Omega_\phi - 2q + 1\,,\qquad
  v^2 \;\leftrightarrow\; 3\Omega_\phi + 2q - 1\,,
\end{equation}
each of which we have checked to be an identity.

%======================================================

\subsection{Coupling curvature
  \texorpdfstring{$\Gamma_F$}{Gamma\_F}}
\label{sec:GammaF}
%------------------------------------------------------
We next reconstruct the coupling
curvature
$\GF = (\dd^2 F/\dd\Phi^2)\,F/(\dd F/\dd\Phi)^2$.
Starting from the slope definition
$\lambda_F = (\dd F/\dd\Phi)/F$ and
applying the chain rule, the curvature
parameter is related to the field
derivative of the slope through
\begin{equation}\label{eq:dlF-dPhi}
  \frac{\dd\lF}{\dd\Phi} = \lF^{\,2}\,(\GF - 1)\,.
\end{equation}
Rewriting the field derivative as
$\dd/\dd\Phi = (1/v)\,\dd/\dln$ and rearranging for $\GF$ gives
\begin{equation}\label{eq:GammaF_form1}
  \GF = 1 + \frac{1}{v\lF^2}
    \frac{\dd\lF}{\dln}\,.
\end{equation}
Differentiating $\lF = \aM/v$ with
respect to $\ln a$ and substituting back
into Eq.~\eqref{eq:GammaF_form1} yields the
compact form
\begin{equation}\label{eq:GammaF_intermediate}
  \GF = 1 + \frac{\aM'}{\aM^2}
    - \frac{v'}{\aM\,v}\,.
\end{equation}
The derivative $v'$ is obtained by
differentiating
Eq.~\eqref{eq:v2_background}, retaining $F'=F\aM$ and using
$q' = q(1+2q) - j$ and
$\Om' = (2q-1)\,\Om$. After algebraic
simplification the full result reads
\begin{equation}\label{eq:GammaF_final}
  \GF = 1 + \frac{\aM'}{\aM^2} - \frac{\mathcal{N}_{F}}{2\,\aM \left[2(1+q)+\aM(2+q-\aM)-\aM'-3\Om \right]} \,,
\end{equation}
where the numerator $\mathcal{N}_F$ is
\begin{equation}\label{eq:GammaF_num}
\begin{split}
  \mathcal{N}_{F} &= (2q^2+q-j)(2+\aM) + \aM'(2+q-3\aM) \\
  &\quad - 3(2q-1)\,\Om + \aM\left[2(1+q)+\aM(2+q-\aM) \right] - \aM{''} \,.
\end{split}
\end{equation}
The coupling curvature $\GF$ draws on the same inputs as $\lU$, the cosmographic parameters
$(q,j)$, the matter density $\Om$, and the Planck-mass running and its first two derivatives $(\aM,\aM',\aM{''})$. It
diverges as $\aM \to 0$, reflecting that $\GF$ is intrinsically ill-defined in
the minimally-coupled limit, where $(\dd F/\dd\Phi)^2$ vanishes. Only the
combination $\lF^2\GF$, which controls the quadratic term of the
reconstructed $F(\Phi)$ in Eq.~\eqref{eq:taylor_F}, remains finite; it is
this product, and not $\GF$ separately, that should be regarded as the
observable of the coupling sector.

%======================================================
\subsection{Potential curvature \texorpdfstring{$\Gamma_U$}{Gamma\_U}}
%======================================================
We construct $\GU$ following the same
procedure used for $\GF$ in
Sec.~\ref{sec:GammaF}. Two new inputs
enter at this order: the snap $s$ from
the cosmographic side and the third
derivative $\aM{'''}$ from the
gravitational side. Differentiating
$\lU$ with respect to $\Phi$ and
rearranging yields
\begin{equation}\label{eq:GammaU_def}
  \GU = 1 - \frac{1}{v\,\lU^2}
    \frac{\dd\lU}{\dln}\,.
\end{equation}
Differentiating $\lU = \NN/(6v\UU)$ with respect to
$\ln a$ and substituting Eq.~\eqref{eq:Uprime} for
$\UU'/\UU$
\begin{equation}
  \frac{\dd\lU}{\dln} = \frac{\NN'}{6v\UU}
    - \lU\frac{v'}{v}
    - 2(1+q)\lU + \lU^2 v\,.
\end{equation}
Substituting into Eq.~\eqref{eq:GammaU_def} then gives

\begin{equation}\label{eq:GammaU_final}
  \GU = \frac{1}{\lU\,v}
    \left[2(1+q) + \frac{v'}{v}
    - \frac{\NN'}{\NN}\right]\,,
\end{equation}
where $v'$ and $\NN'$ follow from
differentiating Eq.~\eqref{eq:v2_background} and
Eq.~\eqref{eq:calN} respectively, and using the cosmographic identities Eq.~\eqref{eq:cosmographic-identities}.
The derivative $\NN'$
takes the  form
\begin{equation}\label{eq:Nprime}
  \NN' = \NN'_0 - (2+\aM)\,s - \aM{'''}
         \;+\; \aM\left(\NN + 9\,\Om\right)\,,
\end{equation}
where the snap $s$ and the third derivative $\aM'''$ enter only through the two
terms shown explicitly, and $\NN_0'$ collects the remaining contributions,
$(q,j,\Om,\aM,\aM',\aM{''})$
\begin{align}\label{eq:N0prime}
  \NN_0' &= \aM'\!\left[(2+\aM)(2{-}\aM) + 3(1{+}\aM)\,q - j\right] \nonumber \\
         &\quad + (2{+}\aM)\!\left[(3q{-}2\aM)\,\aM' + 3(1{+}\aM)(q{+}2q^2{-}j) - j(2{+}3q)\right] \nonumber \\
         &\quad + 3\aM'\left[q{+}2q^2{-}j - \,\aM'\right] \nonumber \\
         &\quad + (3q - 2 - 3\aM)\,\aM{''} - 9(2q{-}1)\,\Om\,.
\end{align}
The final term of Eq.~\eqref{eq:Nprime} deserves comment. The quantity
$\NN$ of Eq.~\eqref{eq:calN} is written at $F=1$, whereas the exact
relation $\lU = \NN/(6v\UU)$ holds for general $F$ with $\NN
\to F\big[(2+2q-\aM)A - A'\big] - 9\Om$ and $A \equiv (2+\aM)(2+\aM-q)+\aM'$.
Differentiating the $F=1$ expression alone therefore omits the contribution of
$F' = F\aM$ acting on the first, $F$-proportional piece, which is exactly
$\aM(\NN+9\Om)$. Retaining it makes Eq.~\eqref{eq:GammaU_final} exact for all
$\aM$; omitting it introduces an error of relative order $\aM(\NN+9\Om)/\NN'$,
which is enhanced over the naive $O(\aM)$ estimate by the large $9\Om$ term,
though still numerically negligible at the LLR scale
(Sec.~\ref{sec:ref_values}).

In the minimally-coupled limit
$\aM = \aM' = \aM{''} = \aM{'''} = 0$, Eqs.~\eqref{eq:GammaU_final}--\eqref{eq:N0prime} collapse to
\begin{equation}\label{eq:GU_mincoupled}
  \GU \;\xrightarrow{\;\aM\to0\;}\;
  \frac{1}{\lU\,v}\left[2(1+q) + \frac{v'}{v} - \frac{\NN'}{\NN}\right],
\end{equation}
with $\lU$ given by Eq.~\eqref{eq:lU_mincoupled} and
\begin{align}
  v^2 &= 2(1+q) - 3\Om\,, \qquad
  \NN = 2\left(4 + 3q - j\right) - 9\,\Om\,, \nonumber\\[4pt]
  \frac{v'}{v} &= \frac{2\left(2q^2+q-j\right) - 3(2q-1)\,\Om}
                       {2\left[2(1+q)-3\Om\right]}\,, \nonumber\\[4pt]
  \NN' &= 6\left(q + 2q^2 - j\right) - 2j(2+3q) - 9(2q-1)\,\Om - 2s\,;
  \label{eq:GU_mincoupled_pieces}
\end{align}
after substituting $\Omega_m = 1 - \Omega_\phi$, and using the dictionary
\eqref{eq:dictionary}, this reduces to Eq.~(12) of Ref.~\cite{CDS2026}. We
have verified the reduction symbolically: the difference of the two
expressions simplifies identically to zero as a rational function of
$(q,j,s,\Omega_\phi)$.

%=====================================================
\subsection{Observable reformulation via the \texorpdfstring{$g_n$}{g\_n}
  tower}\label{sec:obs}
%=====================================================
The gravitational  tower $(\aM,\aM',\aM{''},\aM{'''})$ is not directly
observable. A direct observational handle is provided by the time variation of
the background gravitational constant $G(a) = 1/F(a)$, whose history we
characterise through the logarithmic derivatives
\begin{equation}\label{eq:gn}
  g_n \equiv \left.\frac{\dd^n\ln G}{\dd(\ln a)^n}\right|_{a=a_0},
  \qquad n = 1,2,3,4\,.
\end{equation}
We adopt logarithmic derivatives with respect to $\ln a$ in
Eq.~\eqref{eq:gn} so that the Planck-mass running and its derivatives
map directly onto the $g_n$. Since $G = 1/F$, one has $\ln G = -\ln F$, and
differentiating gives
\begin{equation}\label{eq:g_to_alpha}
  \aM = -g_1\,,\quad \aM' = -g_2\,,\quad \aM'' = -g_3\,,\quad \aM''' = -g_4\,.
\end{equation}
The coefficients $g_n$ are connected to the present-day time derivatives of
$G(t)$. The first four read
\begin{align}
  g_1 &= \frac{\dot G}{G H}\,, \label{eq:g1_obs}\\[2pt]
  g_2 &= \frac{\ddot G}{G H^2} + (1+q)\,g_1 - g_1^2\,, \label{eq:g2_obs}
\end{align}
\begin{align}
g_3 ={}& \frac{\dddot G}{G H^3}
  + 3(1+q)\,\frac{\ddot G}{G H^2}
  - 3\,\frac{\dot G\,\ddot G}{G^2 H^3}
  + \big(1+3q+3q^2-j\big)\frac{\dot G}{G H}\nonumber\\
  &- 3(1+q)\,\frac{\dot G^2}{G^2 H^2}
  + 2\,\frac{\dot G^3}{G^3 H^3}\,, \label{eq:g3_obs}\\[2pt]
g_4 ={}& \frac{\ddddot G}{G H^4}
  + 6(1+q)\,\frac{\dddot G}{G H^3}
  - 4\,\frac{\dot G\,\dddot G}{G^2 H^4}
  - 3\,\frac{\ddot G^2}{G^2 H^4}
  + \big(7+18q+15q^2-4j\big)\frac{\ddot G}{G H^2}\nonumber\\
  &- (18+18q)\,\frac{\dot G\,\ddot G}{G^2 H^3}
  + 12\,\frac{\dot G^2\,\ddot G}{G^3 H^4}
  + \big(1+7q+18q^2+15q^3-6j-10jq-s\big)\frac{\dot G}{G H}\nonumber\\
  &- \big(7+18q+15q^2-4j\big)\frac{\dot G^2}{G^2 H^2}
  + 12(1+q)\,\frac{\dot G^3}{G^3 H^3}
  - 6\,\frac{\dot G^4}{G^4 H^4}\,. \label{eq:g4_obs}
\end{align}
The recursion generating Eqs.~\eqref{eq:g1_obs}--\eqref{eq:g4_obs}, together
with a consistency check, is given in Appendix~\ref{app:gtower}. A convenient
test of any member of the tower is the power-law case $G \propto a^n$, for
which $g_1 = n$ and $g_{n\ge2} = 0$ identically; each of
Eqs.~\eqref{eq:g2_obs}--\eqref{eq:g4_obs} satisfies this.

At first order $g_1 = d\ln G/d\ln a\,|_{a_0} = (\dot G/G)/H_0$
coincides, up to the factor $H_0$, with the present-day fractional
rate $\dot G/G$ probed by LLR.
The second coefficient $g_2$ can be bounded through
theoretical viability arguments and, independently, by large-scale-structure
constraints on the Planck-mass running, as discussed below. The higher
coefficients $g_3$ and $g_4$ involve the third and fourth time
derivatives  $\dddot G$ and $\ddddot G$, which are themselves
unconstrained by present observations; we therefore treat $g_3$ and
$g_4$ as free parameters in the reconstruction.

Substituting Eq.~\eqref{eq:g_to_alpha} into
Eqs.~\eqref{lambda_u},~\eqref{eq:GammaF_final},~\eqref{eq:GammaU_final}
 yields the fully observable forms of all
 reconstructed quantities. The coupling slope follows directly from
 Eq.~\eqref{eq:alphaM} with $\aM = -g_1$,
\begin{equation}\label{eq:lambdaF_obs}
  \lF = \frac{-g_1}{\sqrt{2 - 2g_1 - g_1^2 - 3\,\Om
    + q(2{-}g_1) + g_2}}\,,
\end{equation}
showing explicitly that $\lF$ vanishes in the limit of a constant
gravitational coupling, $g_1 \to 0$.
The potential slope is
 \begin{equation}\label{eq:lambdaU_obs}
  \lU = \frac{(2{-}g_1)\!\left[(2+g_1)(2{-}g_1) + 3(1{-}g_1)\,q - j\right]
    - (3q - 2 + 3g_1)\,g_2 - 9\,\Om + g_3}
    {\sqrt{2 - 2g_1 - g_1^2 - 3\,\Om + q(2{-}g_1) + g_2}\,
      \left[(2{-}g_1)(2{-}g_1{-}q) - g_2 - 3\,\Om\right]}\,,
\end{equation}
which now depends on the cosmographic jerk $j$ and $g_3$ in addition to the
quantities entering $\lF$.
The coupling curvature is
\begin{equation}\label{eq:GammaF_obs}
\begin{split}
  & \quad  \quad\GF = 1 - \frac{g_2}{g_1^{\,2}}+\\
   &\frac{(2q^2 + q - j)(2{-}g_1) - g_2(2 + q + 3g_1) - 2g_1(1+q) + g_1^{2}(2+q+g_1) - 3(2q - 1)\,\Om + g_3}
         {2\,g_1\!\left[2 - 2g_1 - g_1^2 - 3\,\Om + q(2{-}g_1) + g_2\right]}\,.
\end{split}
\end{equation}
The full observable form of $\GU$ is unwieldy; we
therefore present it through its constituent quantities, from which
Eq.~\eqref{eq:GammaU_final} assembles the result. In terms of observables these
read

\begin{align}
  \NN  &= (2-g_1)\!\left[(2+g_1)(2-g_1) +3(1-g_1)q-j\right] \notag\\
  &\quad - (3q-2+3g_1)\,g_2 - 9\,\Om + g_3\,, \label{eq:appN}\\[6pt]
  \frac{v'}{v} &= \frac{(2q^2+q-j)(2-g_1) - g_2(2+q+3g_1)}{2\!\left[2 - 2g_1 - g_1^2 - 3\Om + q(2-g_1) + g_2\right]} \notag\\[6pt]
  &\quad - \frac{g_1[2(1+q)-g_1(2+q+g_1)] + 3(2q-1)\,\Om - g_3}{2\!\left[2 - 2g_1 - g_1^2 - 3\Om + q(2-g_1) + g_2\right]}\,, \label{eq:appVprime}
\end{align}
and $\NN'$ takes the form
\begin{equation}\label{eq:Nprime_obs}
  \NN' = \NN'_0 - (2-g_1)\,s + g_4 - g_1\!\left(\NN + 9\,\Om\right)
\end{equation}
with
\begin{align} \label{eq:appN0prime}
  \NN_0' &= -g_2\!\left[(2-g_1)(2{+}g_1) + 3(1{-}g_1)\,q - j\right] \nonumber \\
         &\quad + (2{-}g_1)\!\left[-(3q{+}2g_1)\,g_2 + 3(1{-}g_1)(q{+}2q^2{-}j) - j(2{+}3q)\right] \nonumber \\
         &\quad - 3\left[q{+}2q^2{-}j + \,g_2\right]g_2 \nonumber \\
         &\quad - (3q - 2 + 3g_1)\,g_3 - 9(2q{-}1)\,\Om\,.  
\end{align}

We have verified Eqs.~\eqref{eq:lambdaF_obs}--\eqref{eq:appN0prime}
symbolically against the defining relations for
$\lU$ and $\GU$, and against their $F$ counterparts,
propagating $q$, $j$, $\Om$ and the $g_n$
with Eq.~\eqref{eq:cosmographic-identities} together with
$g_n' = g_{n+1}$, and retaining $F' = F\aM$ throughout.

Equations~\eqref{eq:lambdaF_obs}--\eqref{eq:appN0prime} make explicit
a natural observational hierarchy. The coupling slope $\lF$ depends
only on $q$, $\Om$, and the first two coefficients $(g_1, g_2)$ of
the gravitational tower; the potential slope $\lU$ and the coupling
curvature $\GF$ additionally require the jerk $j$ and $g_3$; and the
potential curvature $\GU$, at next order, requires the snap $s$ and
the fourth gravitational coefficient $g_4$. Each step up the
reconstruction hierarchy thus requires one additional cosmographic
parameter and one additional gravitational derivative.
The inputs required for each quantity are summarised in
Table~\ref{tab:inputs}.

\begin{table}[ht]
\centering
\begin{tabular}{ll}
\hline\hline
Parameter & Required inputs \\
\hline
$\lF$ & $q,\,\Om,\,g_1,\,g_2$ \\
$\lU$ & $q,\,j,\,\Om,\,g_1,\,g_2,\,g_3$ \\
$\GF$  & $q,\,j,\,\Om,\,g_1,\,g_2,\,g_3$ \\
$\GU$  & $q,\,j,\,s,\,\Om,\,g_1,\,g_2,\,g_3,\,g_4$ \\
\hline\hline
\end{tabular}
\caption{Observational inputs required for each
reconstructed quantity.}
\label{tab:inputs}
\end{table}
%=================================================

\section{Reconstruction of \texorpdfstring{$U(\Phi)$}{U(Phi)} and
  \texorpdfstring{$F(\Phi)$}{F(Phi)}}\label{sec:reconstruction}
%=================================================
Given the slope and curvature parameters
derived in Sec.~\ref{sec:obs}, the two
free functions $U(\Phi)$ and $F(\Phi)$ can
be expanded in Taylor series around the
present-day field value $\Phi_0$. The
reconstruction of $U(\Phi)$ generalises
that of Ref.~\cite{CDS2026} to scalar-tensor
gravity, while $F(\Phi)$ is a new result
with no analogue in the minimally-coupled
case. In what follows
we evaluate both against current cosmographic data.

%------------------------------------------------------

\subsection{Taylor expansion of \texorpdfstring{$U(\Phi)$}{U(Phi)}
  and \texorpdfstring{$F(\Phi)$}{F(Phi)}}\label{sec:taylor}
%-----------------------------------------------------
 Setting $\delta\Phi \equiv \Phi - \Phi_0$, the Taylor
series  of the two functions read
\begin{align}
  U(\Phi) &= U(\Phi_0)
    + \frac{dU}{d\Phi}\bigg|_0 \delta\Phi
    + \frac{1}{2}\frac{d^2U}{d\Phi^2}\bigg|_0 (\delta\Phi)^2
    + \cdots,
  \label{eq:taylor_U_bare}\\[4pt]
  F(\Phi) &= F(\Phi_0)
    + \frac{dF}{d\Phi}\bigg|_0 \delta\Phi
    + \frac{1}{2}\frac{d^2F}{d\Phi^2}\bigg|_0 (\delta\Phi)^2
    + \cdots.
  \label{eq:taylor_F_bare}
\end{align}
Introducing the slope and curvature parameters $\lU,\GU,\lF,\GF$ of
Sec.~\ref{sec:obs} and using the normalisations $U_0 \equiv U(\Phi_0) =
3H_0^2\,\UU$, the present-day potential energy, and $F(\Phi_0)=1$, we obtain
\begin{align}
  U(\Phi) &= U_0\!\left[1 - \lU\,\delta\Phi
    + \tfrac{1}{2}\,\lU^2\,\GU\,(\delta\Phi)^2 + \cdots\right],
  \label{eq:taylor_U}\\[2pt]
  F(\Phi) &= 1 + \lF\,\delta\Phi
    + \tfrac{1}{2}\,\lF^2\,\GF\,(\delta\Phi)^2 + \cdots\,,
  \label{eq:taylor_F}
\end{align}
with $\delta\Phi$ in reduced Planck units. Each order in the expansions draws on the next parameter of each tower,
so the quadratic truncation in
Eqs.~\eqref{eq:taylor_U},\eqref{eq:taylor_F} requires the cosmographic
tower only up to the snap $s$ and the gravitational tower only up to
$g_4$. Extending to cubic order would bring in the fifth cosmographic
parameter and $g_5$. We do not pursue this: as noted in Sec.~\ref{sec:cosmography}, the snap is already only weakly constrained, and the higher coefficients would carry little observational weight.

We plot the reconstructions over $|\delta\Phi| \le 0.1\,\Mpl$. This range is
set by the truncation itself rather than by the data: the ratio of the
quadratic to the linear term in Eq.~\eqref{eq:taylor_U} is
$\tfrac{1}{2}\lU\GU\,\delta\Phi$, which for the largest curvature obtained
below with a finite slope, $\lU\GU \simeq 3$, reaches $\simeq 0.15$ at
$|\delta\Phi| = 0.1$. Beyond this the neglected cubic term is no longer
guaranteed to be subdominant and the expansion should not be trusted.

%-------------------------------------------------
\subsection{Parameter values and viability}\label{sec:ref_values}
%-----------------------------------------------
We adopt the present-day cosmographic values from
Pourojaghi \emph{et al.}~\cite{Pourojaghi:2025}  and
Mishra \emph{et al.}~\cite{Mishra:2026} collected in
Table~\ref{tab:cosmo} spanning supernova-only fits and DESI~DR2
combinations across three expansion methods; the first two rows are the
Pad\'e(3,2) fits of Table~2 of Ref.~\cite{Pourojaghi:2025}, in which $q_0$,
$j_0$ and $s_0$ are marginalised over the two higher cosmographic parameters
$l_0$ and $m_0$, and the remaining three are from Table~1 of
Ref.~\cite{Mishra:2026}. For a wider survey of current cosmographic constraints see Appendix~\ref{app:cosmography_compilation}.
\begin{table}[ht]
\centering
\caption{Cosmographic best-fit values with $1\sigma$ uncertainties.
Top: Pad\'e(3,2) fits of Pourojaghi \emph{et al.}~\cite{Pourojaghi:2025}.
Bottom: Taylor, Pad\'e(2,2) and Chebyshev fits of Mishra
\emph{et al.}~\cite{Mishra:2026} to DESI~DR2. The last column gives the
critical matter density $\Omega_{\rm m}^{\rm crit} = \tfrac{2}{3}(1+q_0)$
of Eq.~\eqref{eq:Omcrit}, above which no real scalar field exists at
$g_n=0$.}
\label{tab:cosmo}

\begin{tabular}{lcccc}
\hline\hline
Dataset & $q_0$ & $j_0$ & $s_0$ & $\Omega_{\rm m}^{\rm crit}$\\
\hline
DES-SN5YR & $-0.503^{+0.043}_{-0.048}$ & $0.97 \pm 0.17$        & $-0.56^{+0.20}_{-0.26}$ & $0.331$\\
Pantheon+ & $-0.465 \pm 0.036$         & $0.85^{+0.19}_{-0.12}$ & $-0.33^{+0.12}_{-0.29}$ & $0.357$\\
\hline\hline
\end{tabular}

\vspace{1.5ex}

\begin{tabular}{lcccc}
\hline\hline
Method (DESI~DR2) & $q_0$ & $j_0$ & $s_0$ & $\Omega_{\rm m}^{\rm crit}$\\
\hline
Taylor ($z \leq 1$) & $-0.41 \pm 0.23$            & $1.48^{+0.96}_{-1.2}$  & $1.64^{+0.43}_{-2.1}$  & $0.393$\\
Pad\'e~$(2,2)$      & $-0.397^{+0.089}_{-0.077}$  & $0.73^{+0.28}_{-0.44}$ & $0.55^{+0.39}_{-1.4}$  & $0.402$\\
Chebyshev           & $-0.472 \pm 0.052$          & $0.59^{+0.16}_{-0.24}$ & $-0.48^{+0.25}_{-0.59}$ & $0.352$\\
\hline\hline
\end{tabular}
\end{table}

We fix $\Omega_{{\rm m},0}=0.3$ throughout, in broad
agreement with independent constraints~\cite{Planck2018,DESIDR2}. This
choice deserves a caveat, since those constraints are themselves derived
assuming a $\Lambda$CDM expansion history and General Relativity, and
importing them into a modified-gravity reconstruction is not strictly
self-consistent. We therefore treat $\Omega_{{\rm m},0}$ as an external prior
rather than a measurement and quantify its impact in
Sec.~\ref{sec:results}. Two features of Table~\ref{tab:cosmo} are worth
noting in this connection. First, the critical density of
Eq.~\eqref{eq:Omcrit} lies as low as $\Omega_{\rm m}^{\rm crit}=0.331$ for
DES-SN5YR, so the fiducial value sits only $10\%$ below the boundary at which
the kinetic term vanishes; the reconstruction is correspondingly sensitive to
$\Omega_{{\rm m},0}$ for that dataset. Second, the constraint runs the other
way as well: for a given $\Omega_{{\rm m},0}$, cosmographies with
$q_0 < \tfrac{3}{2}\Omega_{{\rm m},0} - 1$ admit no quintessence description
at all within this framework.

The leading gravitational coefficient is constrained by LLR, which yields the most precise limits on the time
variation of the gravitational coupling. Reference~\cite{Biskupek2021}
reports two solutions: estimating the linear and quadratic terms separately
gives $\dot G/G = (-5.0\pm9.6)\times10^{-15}\,\mathrm{yr}^{-1}$ and
$\ddot G/G = (1.6\pm2.0)\times10^{-16}\,\mathrm{yr}^{-2}$, while a joint fit,
in which the two are correlated at the $70\%$ level, gives
$\dot G/G = (0.2\pm1.3)\times10^{-14}\,\mathrm{yr}^{-1}$ and
$\ddot G/G = (2.0\pm2.8)\times10^{-16}\,\mathrm{yr}^{-2}$. Since $g_1$ and
$g_2$ enter our expressions simultaneously, the joint solution is the
appropriate one here; both supersede the earlier analysis of
Ref.~\cite{Hofmann2018}. With a
fiducial
$H_0=70\,\mathrm{km\,s^{-1}\,Mpc^{-1}}=7.2\times10^{-11}\,
\mathrm{yr}^{-1}$ the joint solution gives $g_1 = (0.03\pm0.18)\times10^{-3}$,
or, quoting the sum of the central value and the $1\sigma$ error as a bound,
\begin{equation}
  |g_1| \lesssim 2\times10^{-4}. \label{eq:g1_bound}
\end{equation}
With $\lF \simeq -g_1/v$ and $v = O(0.3)$ this gives
$(\dd F/\dd\Phi)^2 = \lF^2 \lesssim 10^{-6}$, so the limit assumed in
Eq.~\eqref{eq:Geff} is satisfied by six orders of magnitude and
$G_{\rm eff}$ and $G$ may indeed be identified.

The second coefficient is far less constrained. Since $g_1$ is of
order $10^{-4}$, the $g_1$-dependent terms in Eq.~\eqref{eq:g2_obs}
are negligible and $g_2$ is dominated by the first term
$\ddot G/(GH^2)$. The measured $\ddot G/G$ then translates into
$g_2 = (3.9\pm5.4)\times10^{4}$, that is $|g_2|\lesssim 10^{5}$. This bound places no meaningful restriction on the
reconstruction.

Far stronger bounds on $g_2$ follow from theoretical
consistency~\cite{Nesseris2007}. Positivity of the scalar-field
kinetic term, $v^2 > 0$ in Eq.~\eqref{eq:v2_background}, requires (for
$|g_1|\ll 1$)
\begin{equation}
  g_2 \;>\; 3\,\Omega_m - 2(1+q_0)\,, \label{eq:g2_viability}
\end{equation}
which for the cosmographic values of Table~\ref{tab:cosmo} gives
$g_2 \gtrsim -0.09$ to $-0.31$ depending on the dataset.  Positivity of the
potential, $\UU>0$ in Eq.~\eqref{eq:U_background}, as required for
the quintessence interpretation, bounds $g_2$ from above,
$g_2 < 2(2-q_0)-3\Omega_m \simeq 3.9$--$4.1$.

An independent handle on $g_2$ is sometimes sought in large-scale structure,
but the two constraints are not straightforwardly compatible and the point
requires care. Cosmological analyses bound the \emph{history} of the
Planck-mass running, usually in the parametrisation $\aM = \hat\alpha_M\,
\Omega_{\rm DE}(a)$, finding $\hat\alpha_M$ of order a few tenths: for
example $\hat\alpha_M = 0.25^{+0.19}_{-0.29}$ from
KiDS+GAMA~\cite{SpurioMancini2019}, with comparable figures from CMB and
redshift-space-distortion combinations~\cite{NollerNicola2019} and from
KiDS-Legacy~\cite{KiDSLegacy2026}. Within that single-amplitude
parametrisation, however, $\aM$ and $\aM'$ are locked together: today
$g_1 = -\hat\alpha_M\Omega_{\rm DE}$ and $g_2 = -\hat\alpha_M\Omega_{\rm DE}'$,
so the central LSS value implies $g_1 \simeq -0.18$, three orders of
magnitude above the LLR bound~\eqref{eq:g1_bound}. Read the other way,
imposing LLR inside the same parametrisation forces
$|g_2| \lesssim 2\times10^{-4}$, which would render the gravitational sector
entirely invisible in the reconstruction. Taken at face value the two
constraints are mutually exclusive, and quoting them side by side as
independent bounds on $g_2$ would be incorrect.

There are two ways out, and the reconstruction developed here commits to one
of them. The LSS analyses cited above are not, in fact, measuring the same
quantity as LLR: they build a Vainshtein screening mechanism into the
non-linear regime by construction, precisely so as to recover general
relativity on small scales~\cite{KiDSLegacy2026}, so their $\hat\alpha_M$ is
a large-scale quantity that is not required to equal the local one. If the
scalar is screened in this way, the tension above is not a contradiction but
a statement that the two probes are complementary; the cost is that
Eq.~\eqref{eq:Geff} then no longer licenses the identification
$G_{\rm eff}\simeq G$, and the mapping of the $g_n$ onto LLR observables that
underpins Sec.~\ref{sec:obs} fails. We therefore adopt the unscreened branch,
in which LLR does constrain the cosmological $g_1$, and the hierarchy
$|g_1| \sim 10^{-4} \ll |g_2| \sim 0.1$ is not a tuning but a statement about
where the field sits: from Eq.~\eqref{eq:alphaM}, $g_1 \to 0$ at fixed
$v \neq 0$ requires $\dd F/\dd\Phi \to 0$, so the coupling function is
stationary at the present field value and $g_2$ measures the curvature at
that stationary point. This is precisely the least-coupling configuration
towards which scalar-tensor cosmologies are driven during matter
domination~\cite{DamourNordtvedt1993a,DamourNordtvedt1993b}, and the
near-parabolic $F(\Phi)$ with an extremum at $\Phi_0$ obtained in
Sec.~\ref{sec:results} is its signature. It should be stated plainly that
this is an assumption of the present analysis rather than a result: the
single-amplitude LSS parametrisation is excluded by it, and a reconstruction
that allowed $\aM$ to be far from its attractor value would be constrained
by LLR to have a negligible gravitational sector throughout.

The higher coefficients $g_3$ and $g_4$ are unconstrained by current
data, so we set $g_3=g_4=0$. Both enter additively: $g_3$
shifts the potential slope by $\delta\lU = g_3/(6v\,\UU) \approx
0.5$--$0.8\,g_3$ across the datasets of Table~\ref{tab:cosmo}, and
$g_4$ enters $\GU$ through $\NN'$ in Eq.~\eqref{eq:Nprime_obs} with unit
coefficient. Values of order unity for either would therefore change the
reconstruction at the same level as the cosmographic uncertainties
themselves.

%-------------------------------------------------
\subsection{Reconstruction results}\label{sec:results}
%------------------------------------------------

\begin{figure}[ht!]
\centering
\includegraphics[width=\textwidth]{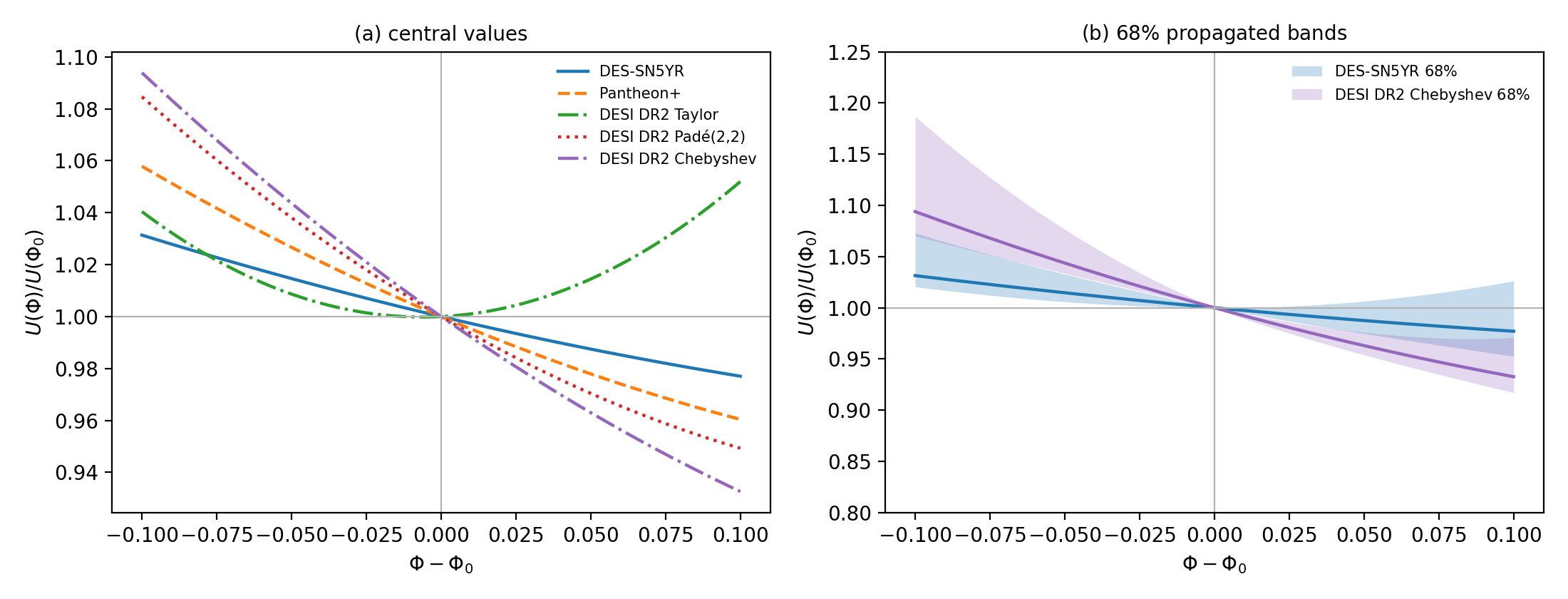}
\caption{Reconstructed potential $U(\Phi)/U(\Phi_0)$ in the
minimally-coupled limit $g_n=0$, for the cosmographic values of
Table~\ref{tab:cosmo}. Left: central values. Right: $68\%$ bands obtained
by propagating the quoted $1\sigma$ cosmographic uncertainties through
Eqs.~\eqref{eq:lambdaU_obs} and~\eqref{eq:GammaU_final}, shown for two
representative datasets, with the central curves overlaid.}
\label{fig:potential_U}
\end{figure}

We first set $g_n=0$, switching off the non-minimal coupling, and verify
that the reconstruction reduces to the minimally-coupled case.
Evaluated at the cosmographic central values of Table~\ref{tab:cosmo},
both the reconstructed parameters (Table~\ref{tab:lambda_gamma}) and the
resulting potentials (Fig.~\ref{fig:potential_U}, left) reproduce those of
Ref.~\cite{CDS2026}.

Central values alone, however, considerably overstate what the data
determine. We therefore propagate the quoted cosmographic uncertainties by
Monte Carlo, sampling $(q_0,j_0,s_0)$ from two-piece Gaussians matched to the
asymmetric errors of Table~\ref{tab:cosmo} and retaining only samples with
$v^2>0$ and $\UU>0$. The results are collected in
Table~\ref{tab:lambda_gamma} and displayed as bands in
Fig.~\ref{fig:potential_U}. Four features of this procedure should be stated
explicitly, since they bound how the intervals may be read.

First, the quoted ranges are \emph{conditional} distributions,
$p(\lambda_U \,|\, v^2>0)$, not marginal posteriors. The conditioning is not
negligible --- it removes between $1\%$ and $27\%$ of the prior volume
depending on the dataset --- and by Eq.~\eqref{eq:v2_w} it is exactly the
removal of the phantom region, discussed below.

Second, the published covariances between $q_0$, $j_0$ and $s_0$ are not
available to us and we sample them independently. This matters at the level
of tens of percent, and not in a direction that can be assumed. Because
$\partial\NN/\partial q = 6$ and $\partial\NN/\partial j = -2$ carry opposite
signs, a positive correlation produces cancellation and narrows the interval,
while a negative correlation does almost nothing. For DES-SN5YR the $68\%$
width of $\lambda_U$ is $0.573$, $0.566$ and $0.341$ for
$\rho(q_0,j_0) = -0.9$, $0$ and $+0.9$ respectively. The independent-sampling
intervals are thus conservative if $q_0$ and $j_0$ are positively correlated,
as is typical of these fits, and essentially unchanged otherwise; they should
not be quoted as formal posteriors.

Third, the uncertainty on $\Omega_{{\rm m},0}$ is not the issue. Floating it
as $0.300\pm0.005$ changes every interval width by less than $1\%$. What
matters is the $\Lambda$CDM-derived central value, which is a systematic and
is quantified separately in Table~\ref{tab:Om_sensitivity}.

Fourth, all quoted errors are conditional on $g_3 = g_4 = 0$. Since $g_3$
shifts $\lambda_U$ by $\delta\lU \approx 0.5$--$0.8\,g_3$, an $O(1)$ value of
$g_3$ would displace $\lambda_U$ by as much as the entire statistical width.
The gravitational sector carries no propagated uncertainty anywhere in this
work; it is explored through benchmarks only.

\begin{table}[ht]
\centering
\caption{Reconstructed potential slope $\lambda_U$ and curvature
$\Gamma_U$ in the minimally-coupled limit $g_n=0$, for
$\Omega_{{\rm m},0}=0.3$ and the cosmographic values of Table~\ref{tab:cosmo}.
Central values are evaluated at the cosmographic best fits; the bracketed
ranges are $68\%$ intervals of the \emph{conditional} distribution
$p(\,\cdot\,|\,v^2>0)$ from the Monte Carlo described in the text, obtained
by sampling $q_0$, $j_0$ and $s_0$ independently and holding
$g_3=g_4=0$. The last column gives the fraction of samples removed by that
conditioning, which by Eq.~\eqref{eq:v2_w} is the fraction with
$w_{\rm eff,0}<-1$.}
\label{tab:lambda_gamma}
\begin{tabular}{lcccc}
\hline\hline
Dataset & $\lambda_U$ & $68\%$ range & $\Gamma_U$ & unphysical \\
\hline
DES-SN5YR            & $0.2717$  & $[+0.01,\,+0.58]$ & $11.39$              & $16.4\%$ \\
Pantheon+            & $0.4875$  & $[+0.26,\,+0.65]$ & $7.69$               & $0.9\%$  \\
DESI DR2 Taylor      & $-0.0579$ & $[-0.85,\,+1.26]$ & $2.76\times10^{3}$   & $27.2\%$ \\
DESI DR2 Pad\'e(2,2) & $0.6769$  & $[+0.42,\,+1.14]$ & $7.41$               & $2.4\%$  \\
DESI DR2 Chebyshev   & $0.8064$  & $[+0.64,\,+1.20]$ & $4.08$               & $6.7\%$  \\
\hline\hline
\end{tabular}
\end{table}

Three points follow. The pattern behind them is the sensitivity relation of
Ref.~\cite{CDS2026}, $\delta\lU/\lU = -2\,\delta j/\NN$, which our
Eq.~\eqref{eq:appN} reproduces with $\partial\NN/\partial j = -(2-g_1)$: the
fractional error on the slope is set by the jerk uncertainty divided by
$\NN$, and therefore blows up wherever the background approaches
$\Lambda$CDM. First, $\lambda_U$ is determined to within a few tenths
for the tighter datasets, and all of them except DESI~DR2 Taylor prefer
$\lambda_U > 0$, that is a potential decreasing in the direction of field
evolution. Second, $\Gamma_U$ is not usefully constrained by any current
dataset: its $68\%$ intervals span one to two orders of magnitude, with upper
limits reaching several hundred, so the quadratic term in
Eq.~\eqref{eq:taylor_U} is at present a benchmark rather than a measurement.
The large central value for DESI~DR2 Taylor is not a feature of the data but
an artefact of $\lambda_U \to 0$ there: $\Gamma_U$ is defined with $\lambda_U^2$
in the denominator, and only the product $\lambda_U^2\Gamma_U$, which stays
finite and modest, enters the reconstruction. Third, and most consequentially,
a substantial fraction of the cosmographically allowed parameter space admits
no minimally-coupled quintessence description at all at
$\Omega_{{\rm m},0}=0.3$: $16\%$ of the DES-SN5YR samples and $27\%$ of the
DESI~DR2 Taylor samples violate $v^2>0$ through Eq.~\eqref{eq:Omcrit}.

\begin{figure}[ht!]
  \centering
  \includegraphics[width=0.72\textwidth]{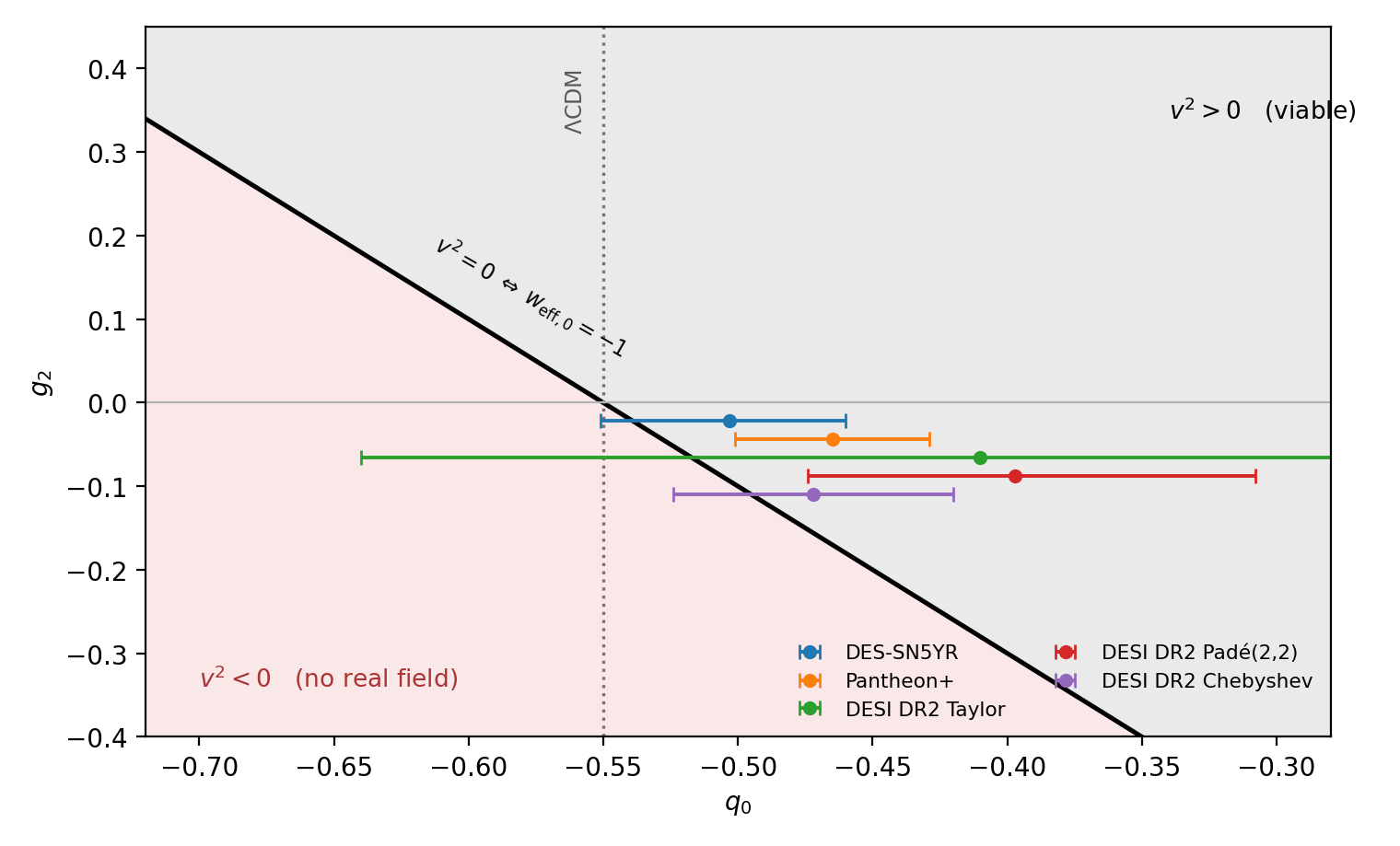}
  \caption{Viability of the scalar-field description in the $(q_0,g_2)$ plane
at $\Omega_{{\rm m},0}=0.3$ and $|g_1|\ll1$. The diagonal is
$v^2 = 2(1+q_0)-3\Om+g_2 = 0$, which by Eq.~\eqref{eq:v2_w} is the locus
$w_{\rm eff,0}=-1$; below it no real scalar field exists. Points with
$1\sigma$ bars are the datasets of Table~\ref{tab:cosmo}, offset vertically
for legibility. A positive $g_2$ admits phantom backgrounds
($q_0 < q_0^{\Lambda}$, dotted line) without introducing a ghost.}
  \label{fig:viability}
\end{figure}

By Eq.~\eqref{eq:v2_w} this excluded region is exactly the set of
cosmographies with $w_{\rm eff,0}<-1$, so the numbers above are
$P(w_{\rm eff,0}<-1)$ for each dataset. The central values all sit on the
non-phantom side --- $w_{\rm eff,0}$ ranges from $-0.955$ for DES-SN5YR to
$-0.854$ for DESI~DR2 Pad\'e(2,2) --- but the $1\sigma$ ranges cross the
divide. This is where the non-minimal coupling does genuine work.
Figure~\ref{fig:viability} shows the boundary in the $(q_0,g_2)$ plane: the
viable region is $g_2 > 3\Om - 2(1+q_0)$, and switching on a positive $g_2$
moves the boundary to the left, admitting phantom backgrounds while keeping
$v^2>0$ and hence avoiding a ghost. The amount required is modest. To
accommodate $q_0$ at its $-1\sigma$ value one needs $g_2 > 0.18$ for
DESI~DR2 Taylor and $g_2 > 0.002$ for DES-SN5YR, while Pantheon+,
Pad\'e(2,2) and Chebyshev need no coupling at all ($g_2 > -0.098$, $-0.152$
and $-0.052$ respectively, all satisfied at $g_2=0$). A cosmographic
determination of $q_0$ below the $\Lambda$CDM value
$q_0^{\Lambda} = \tfrac32\Omega_{{\rm m},0}-1$ would therefore not rule out a
scalar-field description; it would instead place a \emph{lower} bound on the
second coefficient of the gravitational tower.

The $\Omega_{{\rm m},0}$ dependence is shown in
Table~\ref{tab:Om_sensitivity}. For the datasets with $q_0$ comfortably
below the critical value the reconstruction is stable at the few-percent
level, but for DES-SN5YR, whose $\Omega_{\rm m}^{\rm crit}=0.331$ is close to
the fiducial choice, $\lambda_U$ moves by $20\%$ over the plausible prior
range and $\Gamma_U$ by considerably more. This sensitivity should be borne
in mind when interpreting supernova-only reconstructions.

\begin{table}[ht]
\centering
\caption{Sensitivity of the reconstructed slope $\lambda_U$ to the assumed
matter density, at $g_n=0$ and the cosmographic central values of
Table~\ref{tab:cosmo}.}
\label{tab:Om_sensitivity}
\begin{tabular}{lccc}
\hline\hline
Dataset & $\Omega_{{\rm m},0}=0.28$ & $0.30$ & $0.32$ \\
\hline
DES-SN5YR            & $0.319$  & $0.272$  & $0.217$  \\
Pantheon+            & $0.505$  & $0.487$  & $0.478$  \\
DESI DR2 Taylor      & $0.026$  & $-0.058$ & $-0.166$ \\
DESI DR2 Pad\'e(2,2) & $0.685$  & $0.677$  & $0.672$  \\
DESI DR2 Chebyshev   & $0.770$  & $0.806$  & $0.898$  \\
\hline\hline
\end{tabular}
\end{table}

\begin{figure}[ht!]
  \centering
  \includegraphics[width=\textwidth]{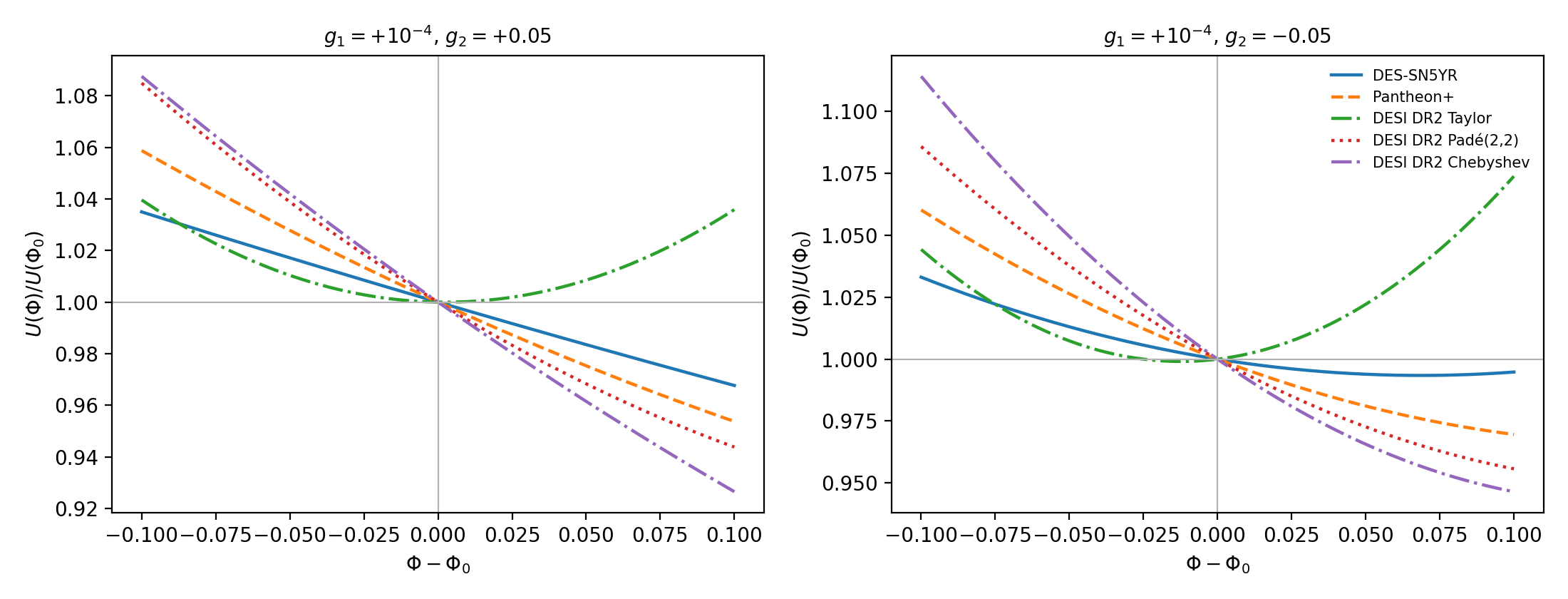}
  \caption{Reconstructed potential $U(\Phi)/U(\Phi_0)$ for $g_1=+10^{-4}$
and $g_2=\pm0.05$, with $g_3=g_4=0$ and $\Omega_{{\rm m},0}=0.3$, for the
cosmographic values of Table~\ref{tab:cosmo}.}
  \label{fig:U_signs}
\end{figure}

We now switch on the non-minimal coupling, taking $g_1 = +10^{-4}$ at
the LLR scale and $g_2 = \pm0.05$, with $g_3 = g_4 = 0$; the sign of
$g_1$ has no visible effect at this magnitude. Among the gravitational
coefficients only $g_2$ matters at the level of the figures. For
$g_2 = +0.05$ the curves stay
close to the minimally-coupled baseline; for $g_2 = -0.05$ the curvature
grows, with DES-SN5YR and DESI~DR2 Taylor developing a minimum near
$\Phi_0$ while the other datasets keep a decreasing slope
(Fig.~\ref{fig:U_signs}).

\begin{figure}[ht!]
  \centering
  \includegraphics[width=\textwidth]{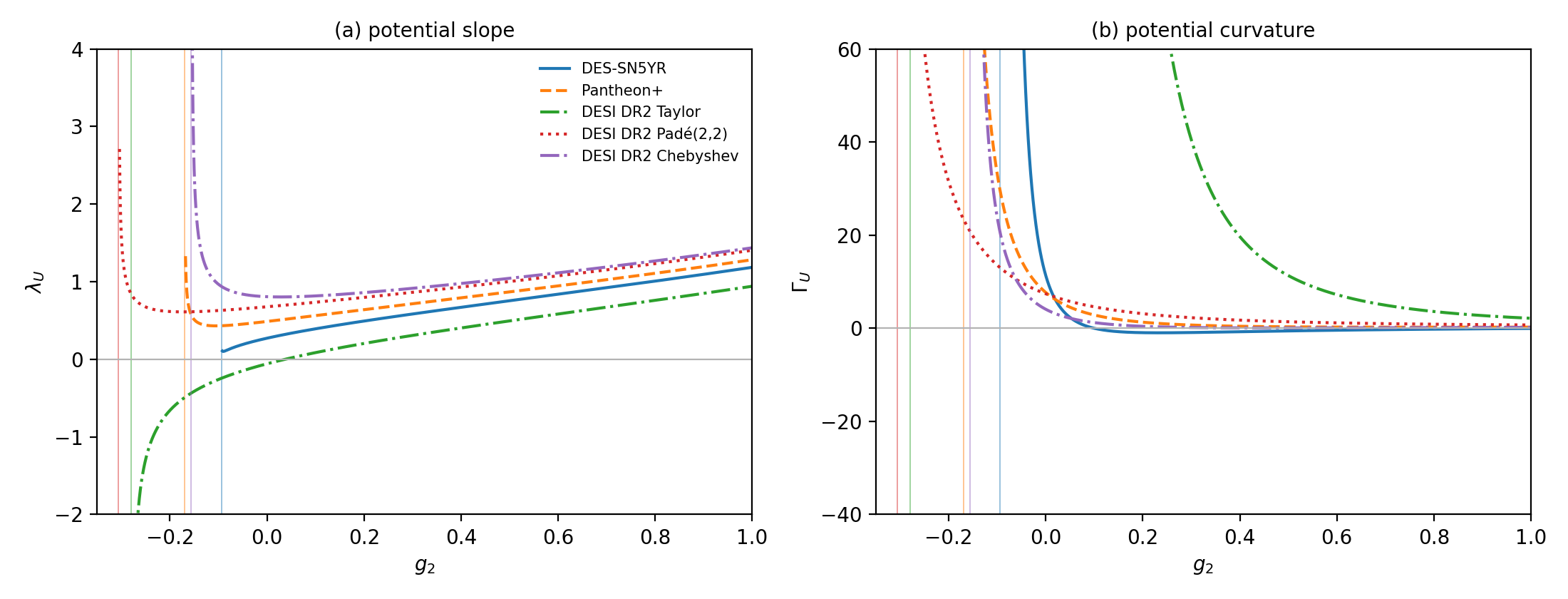}
  \caption{Dependence of the reconstructed potential slope (left) and
curvature (right) on the second gravitational coefficient $g_2$, at
$g_1=+10^{-4}$, $g_3=g_4=0$ and $\Omega_{{\rm m},0}=0.3$. Vertical lines mark
the kinetic-term boundary $g_2 = 3\Om - 2(1+q_0)$ of
Eq.~\eqref{eq:g2_viability} for each dataset, at which $v^2\to0$ and both
quantities diverge.}
  \label{fig:g2_scan}
\end{figure}

Our choice of $|g_2|=0.05$, rather than a value nearer the viability floor,
is deliberate. As $g_2$ approaches the bound of
Eq.~\eqref{eq:g2_viability} the kinetic term vanishes and both $\lU \propto
\NN/v$ and $\GU$ diverge, so any apparent enhancement of the reconstructed
curvature there reflects $v \to 0$ rather than a physical response of the
potential to the coupling. For DES-SN5YR the floor is $g_2 > -0.094$, so
$g_2=-0.09$ would give $v^2 = 0.004$, a factor of $20$ below its
minimally-coupled value, and a correspondingly inflated $\GU$. Figure
\ref{fig:g2_scan} shows the full dependence of $\lU$ and $\GU$ on $g_2$
across the viability window, with the kinetic boundary of each dataset
marked; the reconstruction is well behaved except within roughly $0.02$ of
the boundary.

\begin{figure}[ht!]
  \centering
  \includegraphics[width=\textwidth]{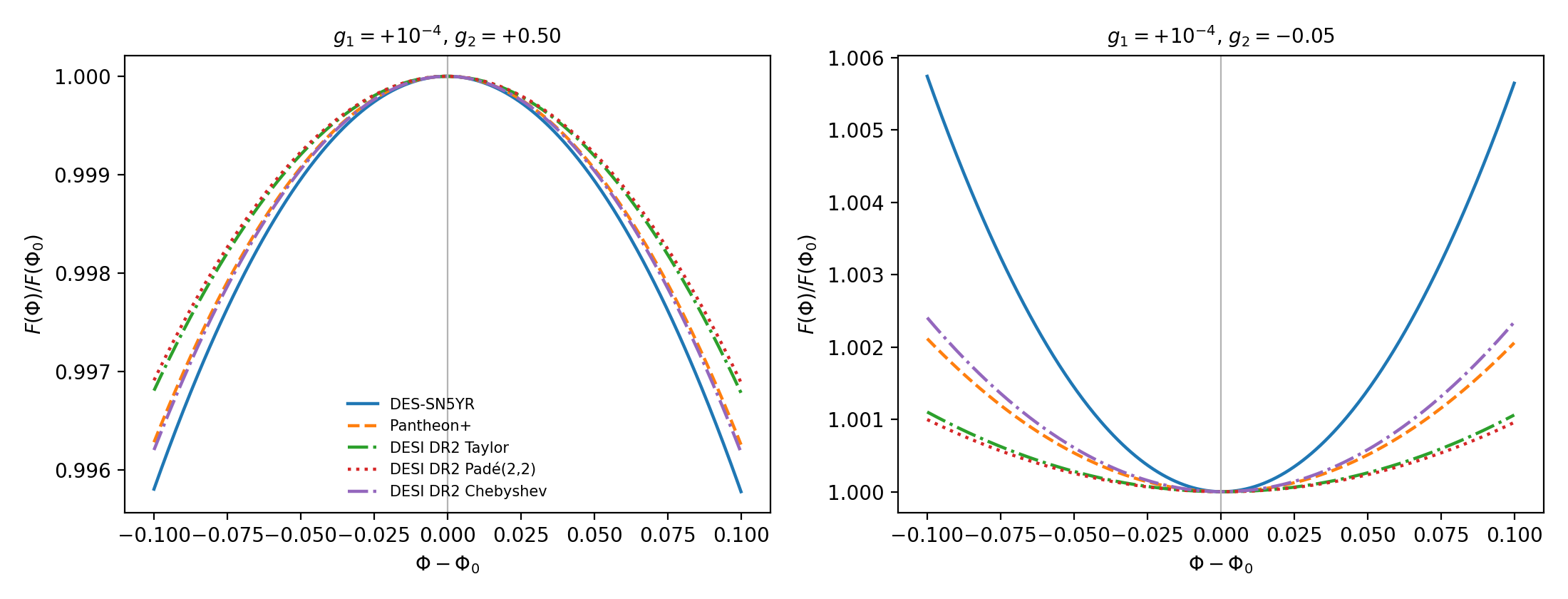}
  \caption{Reconstructed coupling $F(\Phi)/F(\Phi_0)$ for $g_1=+10^{-4}$
and $g_2=+0.50$ (left), $-0.05$ (right), with $\Omega_{{\rm m},0}=0.3$ and the
cosmographic values of Table~\ref{tab:cosmo}.}
  \label{fig:F_g1g2}
\end{figure}

Turning to the coupling sector, Table~\ref{tab:F_sector} lists $\lF$, $\GF$
and their product for the two benchmarks. As anticipated in
Sec.~\ref{sec:GammaF}, neither $\lF \sim -10^{-4}$ nor $\GF \sim 10^{6}$--$10^{7}$
is individually meaningful; only the combination
$\lF^2\GF$, reproduced in the last column by $-g_2/v^2$ to better than
$0.5\%$, is physical. Since the linear term is negligible,
$F(\Phi)$ is a parabola with quadratic coefficient
$\lambda_F^2\GF \simeq -g_2/v^2$, and the sign of $g_2$ fixes its concavity.
The curves nearly coincide
for $g_2=+0.50$, where the $v^2$ are large and $1/v^2$ varies little
across datasets, but fan apart for $g_2=-0.05$, where $v^2$ is both smaller
and more dataset-dependent.
We take these two values as representative of the two
branches of the viability window
(Fig.~\ref{fig:F_g1g2}). Note that the total variation of $F$ over the
plotted range is below $0.5\%$: the LLR bound on $g_1$ forces the coupling
to be nearly flat, and $g_2$ controls only its curvature.

\begin{table}[ht]
\centering
\caption{Coupling-sector quantities at $g_1=+10^{-4}$, $g_3=0$ and
$\Omega_{{\rm m},0}=0.3$. Neither $\lambda_F$ nor $\Gamma_F$ is separately
meaningful; the physical combination is $\lambda_F^2\Gamma_F$, which is
reproduced by $-g_2/v^2$.}
\label{tab:F_sector}
\begin{tabular}{lccccc}
\hline\hline
Dataset & $g_2$ & $v^2$ & $\lambda_F$ & $\Gamma_F$ & $\lambda_F^2\Gamma_F$ \\
\hline
DES-SN5YR            & $+0.50$ & $0.594$ & $-1.30\times10^{-4}$ & $-5.00\times10^{7}$ & $-0.842$ \\
Pantheon+            & $+0.50$ & $0.670$ & $-1.22\times10^{-4}$ & $-5.00\times10^{7}$ & $-0.747$ \\
DESI DR2 Taylor      & $+0.50$ & $0.780$ & $-1.13\times10^{-4}$ & $-5.00\times10^{7}$ & $-0.641$ \\
DESI DR2 Pad\'e(2,2) & $+0.50$ & $0.806$ & $-1.11\times10^{-4}$ & $-5.00\times10^{7}$ & $-0.621$ \\
DESI DR2 Chebyshev   & $+0.50$ & $0.656$ & $-1.24\times10^{-4}$ & $-5.00\times10^{7}$ & $-0.762$ \\
\hline
DES-SN5YR            & $-0.05$ & $0.044$ & $-4.78\times10^{-4}$ & $+4.99\times10^{6}$ & $+1.139$ \\
Pantheon+            & $-0.05$ & $0.120$ & $-2.89\times10^{-4}$ & $+5.00\times10^{6}$ & $+0.417$ \\
DESI DR2 Taylor      & $-0.05$ & $0.230$ & $-2.09\times10^{-4}$ & $+4.97\times10^{6}$ & $+0.216$ \\
DESI DR2 Pad\'e(2,2) & $-0.05$ & $0.256$ & $-1.98\times10^{-4}$ & $+5.00\times10^{6}$ & $+0.196$ \\
DESI DR2 Chebyshev   & $-0.05$ & $0.106$ & $-3.07\times10^{-4}$ & $+5.03\times10^{6}$ & $+0.475$ \\
\hline\hline
\end{tabular}
\end{table}

Finally, we note the numerical size of the correction term in
Eq.~\eqref{eq:Nprime}. At $g_1 = 10^{-4}$ it shifts $\NN'$ by a relative
$1.4\times10^{-3}$ and $\GU$ by $2\times10^{-5}$ to $10^{-3}$ depending on the
dataset, far below the uncertainties of Table~\ref{tab:lambda_gamma}. It is
therefore irrelevant for the present data, and we retain it only so that
Eqs.~\eqref{eq:GammaU_final} and~\eqref{eq:Nprime_obs} are exact for arbitrary
$\aM$, as would be required were the framework applied to a regime with
larger Planck-mass running.

\clearpage
%=========================================

\section{Summary}
\label{sec:summary}
We have generalised the cosmographic reconstruction of
Ref.~\cite{CDS2026} to scalar-tensor gravity, deriving exact
closed-form expressions for the potential slope $\lambda_U$, the
coupling slope $\lambda_F$, and their curvatures $\Gamma_U$ and
$\Gamma_F$ in terms of the cosmographic parameters $(q,j,s)$, the
matter density $\Omega_m$, and the Planck-mass running $\alpha_M$. The
new feature of the scalar-tensor case is the time variation
of the gravitational coupling, which we described through the
gravitational tower $g_n \equiv d^n\ln G/d(\ln a)^n|_{a_0}$. This maps
the theoretical $\alpha_M$ tower onto the observable $\dot G/G$ and its
higher time derivatives, so that all four quantities are expressed
directly in measurable terms. The reconstruction follows a natural
hierarchy, each successive quantity requiring one further parameter
from each tower. In the minimally-coupled limit $g_n=0$ the expressions
reduce to Eqs.~(11) and~(12) of Ref.~\cite{CDS2026}; we verified both
reductions symbolically and reproduced that reference's tabulated
$(\lambda_0,\Gamma_0)$ and Taylor coefficients for all five datasets.

We evaluated the reconstruction against current DESI~DR2, supernova,
and LLR data. The leading coefficient is bounded at the LLR scale,
$|g_1|\lesssim2\times10^{-4}$, while the higher coefficients, unconstrained by
present data, were varied as benchmarks. Setting $g_n=0$ recovers the
slopes, curvatures, and potentials of Ref.~\cite{CDS2026}. Once the
coupling is switched on, the potential $U(\Phi)$ responds only weakly,
through $g_2$, whereas the coupling $F(\Phi)$ is shaped entirely by the
gravitational sector, making $F(\Phi)$ far more
sensitive to $g_2$ than $U(\Phi)$. The near-flat, extremal $F(\Phi)$ that
results is the configuration expected from the Damour--Nordtvedt
least-coupling attractor rather than a tuning.

Propagating the cosmographic uncertainties changes the complexion of the
results. The potential slope $\lambda_U$ is stable across
datasets and determined to within a few tenths, but $\Gamma_U$ is not
usefully constrained by any current dataset, and can in addition be
amplified sharply where $\lambda_U$ approaches zero, driven by the
poorly-constrained jerk and depending further on the weakly-constrained
snap. These intervals are conditional on $v^2>0$, on $g_3=g_4=0$, and on
sampling $q_0$, $j_0$ and $s_0$ independently; the last of these makes them
conservative by up to $40\%$ if the cosmographic parameters are positively
correlated, as is typical.

The clearest new statement concerns the kinematic bound
$\Omega_{\rm m} < \tfrac23(1+q_0)$. Because $v^2 = 3\Omega_\phi(1+w)$
identically, this is not a technical requirement but the condition
$w_{\rm eff,0}\ge-1$, and the $1\%$ to $27\%$ of parameter space it removes
is exactly the phantom region. The non-minimal coupling reopens that region
through $g_2$ alone, requiring $g_2 > 0.18$ at the $-1\sigma$ value of $q_0$
for DESI~DR2 Taylor and less for every other dataset. A cosmographic
determination of $q_0$ below the $\Lambda$CDM value would thus not exclude a
scalar-field description but would set a lower bound on the gravitational
sector. Improving the determination of $q_0$ and $j_0$, rather than
tightening $\dot G/G$, is what would most sharpen this class of
reconstruction.

%====================================================
\section*{Acknowledgements}
This research was supported by COST Action CA21136 -- Addressing
observational tensions in cosmology with systematics and fundamental physics
(CosmoVerse), supported by COST (European Cooperation in Science and
Technology).

\section*{Data availability}
No new observational data were generated in support of this work. The
cosmographic parameters used here are taken from Table~2 of
Ref.~\cite{Pourojaghi:2025} and Table~1 of Ref.~\cite{Mishra:2026}, and the
Lunar Laser Ranging constraints from Ref.~\cite{Biskupek2021}. The code that
reproduces every figure, table and quoted number in this paper, together with
the symbolic verification of the derivation and of its minimally-coupled
limit, is publicly available at \url{https://github.com/leandros11/st-reconstruction}.

%====================================================
\bibliographystyle{unsrt}
\bibliography{references}
\clearpage

\clearpage
\appendix
\section{Compilation of recent cosmographic constraints}
\label{app:cosmography_compilation}
We collect in Table~\ref{tab:cosmography_literature} recent
model-independent constraints on the present-day cosmographic
parameters $(q_0, j_0, s_0)$, restricted to analyses in which these
parameters are fitted directly to data as free parameters of a
kinematic expansion or non-parametric reconstruction.

\begin{table}[ht]
\centering
\footnotesize
\setlength{\tabcolsep}{3.5pt}
\caption{Recent model-independent constraints of the present-day
cosmographic parameters. All analyses assume a spatially flat FLRW
background unless otherwise noted. Errors are $1\sigma$ unless
otherwise indicated. The first
row gives the flat $\Lambda$CDM prediction for $\Omega_{{\rm m},0}=0.3$.}
\label{tab:cosmography_literature}
\begin{tabular}{llllll}
\toprule
Ref. & Data & Method & $q_0$ & $j_0$ & $s_0$ \\
\midrule
\multicolumn{6}{l}{\textit{Reference value}}\\
--- & $\Lambda$CDM, $\Omega_{{\rm m},0}=0.3$ &  & $-0.55$ & $1$ & $-0.35$ \\
\midrule
\multicolumn{6}{l}{\textit{Type Ia supernovae }}\\
\cite{Pourojaghi:2025} & DES-SN5YR & Pad\'e (3,2) & $-0.503^{+0.043}_{-0.048}$ & $0.97 \pm 0.17$ & $-0.56^{+0.20}_{-0.26}$ \\
\cite{Pourojaghi:2025} & Pantheon+ & Pad\'e (3,2) & $-0.465 \pm 0.036$ & $0.85^{+0.19}_{-0.12}$ & $-0.33^{+0.12}_{-0.29}$ \\
\multirow{3}{*}{\cite{Hu:2024}} & Pantheon+ & Pad\'e (2,1) & $-0.35^{+0.08}_{-0.07}$ & $0.43^{+0.38}_{-0.56}$ & --- \\
 & Pantheon+ & Pad\'e (2,2) & $-0.33 \pm 0.09$ & $0.13^{+0.94}_{-0.68}$ & $-0.56^{+2.92}_{-0.98}$ \\
 & Pantheon+ & Pad\'e (3,2) & $-0.22 \pm 0.09$ & $-1.58^{+1.10}_{-0.78}$ & --- \\
\midrule
\multicolumn{6}{l}{\textit{DESI DR1 BAO }}\\
\cite{Pourojaghi:2025} & DR1 & Pad\'e (3,2) & $-0.669^{+0.088}_{-0.076}$ & $1.44^{+0.12}_{-0.20}$ & $0.34 \pm 0.21$ \\
\multirow{3}{*}{\cite{Luongo:2024}} & DR1$+$OHD & Taylor ($z$) & $-0.44^{+0.10}_{-0.16}$ & $0.54^{+0.17}_{-0.10}$ & $-0.48^{+0.24}_{-0.27}$ \\
 & DR1$+$SNe & Taylor ($z$) & $-0.48^{+0.05}_{-0.09}$ & $0.58^{+0.13}_{-0.10}$ & $-0.47^{+0.20}_{-0.22}$ \\
 & DR1$+$OHD$+$SNe & Taylor ($z$) & $-0.50^{+0.06}_{-0.08}$ & $0.57^{+0.14}_{-0.07}$ & $-0.49^{+0.14}_{-0.21}$ \\
\midrule
\multicolumn{6}{l}{\textit{DESI DR2 BAO }}\\
\multirow{4}{*}{\cite{Mishra:2026}} & DR2 & Taylor ($z$) & $-0.41 \pm 0.23$ & $1.48^{+0.96}_{-1.2}$ & $1.64^{+0.43}_{-2.1}$ \\
 & DR2 & Pad\'e (2,1) & $-0.49^{+0.12}_{-0.10}$ & $1.27^{+0.37}_{-0.61}$ & --- \\
 & DR2 & Pad\'e (2,2) & $-0.397^{+0.089}_{-0.077}$ & $0.73^{+0.28}_{-0.44}$ & $0.55^{+0.39}_{-1.4}$ \\
 & DR2 & Chebyshev & $-0.472 \pm 0.052$ & $0.59^{+0.16}_{-0.24}$ & $-0.48^{+0.25}_{-0.59}$ \\
\multirow{4}{*}{\cite{Rodrigues:2025}$^{a}$} & DR2 & Taylor ($y$) & $-0.42^{+0.17}_{-0.13}$ & $1.77^{+0.61}_{-0.86}$ & $>2.63$ \\
 & DR2$+$Union3 & Taylor ($y$) & $-0.47^{+0.10}_{-0.086}$ & $1.93^{+0.41}_{-0.58}$ & $>4.57$ \\
 & DR2$+$Pantheon+ & Taylor ($y$) & $-0.48^{+0.079}_{-0.069}$ & $1.96^{+0.37}_{-0.50}$ & $>5.01$ \\
 & DR2$+$DESY5 & Taylor ($y$) & $-0.51^{+0.068}_{-0.060}$ & $2.04^{+0.30}_{-0.42}$ & $>5.93$ \\
\midrule
\multicolumn{6}{l}{\textit{DESI DR2 $+$ strong lensing }}\\
\multirow{3}{*}{\cite{Kumar:2025a,Kumar:2025b}$^{b}$} & SGL$+$Pantheon+$+$DR2 & Taylor ($y$) & $-0.477^{+0.054}_{-0.058}$ & $0.835^{+0.427}_{-0.394}$ & $0.629^{+2.249}_{-1.786}$ \\
 & SGL$+$Union3$+$DR2 & Taylor ($y$) & $-0.350 \pm 0.077$ & $0.079^{+0.482}_{-0.453}$ & $-2.262^{+1.775}_{-1.128}$ \\
 & SGL$+$DESY5$+$DR2 & Taylor ($y$) & $-0.546^{+0.087}_{-0.081}$ & $1.344^{+0.628}_{-0.616}$ & $3.335^{+3.951}_{-3.191}$ \\
\midrule
\multicolumn{6}{l}{\textit{Non-parametric reconstructions}}\\
\cite{Liu:2023} & CC$+$BAO$+$Pantheon+ & Gaussian proc. & $-0.71 \pm 0.04$ & $1.26 \pm 0.18$ & $0.04 \pm 0.59$ \\
\cite{Jesus:2024} & $H(z)$$+$SNe & Gaussian proc. & $-0.60^{+0.21}_{-0.18}$ & $0.90^{+0.75}_{-0.65}$ & $-0.57^{+0.52}_{-0.31}$ \\
\bottomrule
\multicolumn{6}{p{0.97\textwidth}}{\footnotesize
$^{a}$ Errors at $2\sigma$; $s_0$ values are $2\sigma$ lower limits.
$^{b}$ Spatial curvature left free rather than fixed to zero.}\\
\end{tabular}
\end{table}

%=========================================
\section{The gravitational tower}
\label{app:gtower}

Equations~\eqref{eq:g1_obs}--\eqref{eq:g4_obs} follow from a two-step
construction which we record here, both to make the pattern explicit and
because the coefficients are easy to get wrong.

Write $u \equiv \ln G$ and let $M_k \equiv G^{(k)}/G$ denote the
$k$-th time derivative of $G$ normalised by $G$. The time derivatives of
$u$ are then the cumulants associated with the moments $M_k$,
\begin{align}
  \dot u        &= M_1\,, \nonumber\\
  \ddot u       &= M_2 - M_1^2\,, \nonumber\\
  \dddot u      &= M_3 - 3M_1M_2 + 2M_1^3\,, \nonumber\\
  \ddddot u     &= M_4 - 4M_1M_3 - 3M_2^2 + 12M_1^2 M_2 - 6M_1^4\,.
  \label{eq:cumulants}
\end{align}
It is the $+12M_1^2M_2$ term of the fourth cumulant that generates the
$+12\,\dot G^2\ddot G/(G^3H^4)$ contribution to $g_4$ in
Eq.~\eqref{eq:g4_obs}.

The $g_n$ are then obtained by repeated application of the operator
$D \equiv H^{-1}\dd/\dd t$, so that $g_n = D^n u$. Since $D$ does not
commute with $H$, each application generates cosmographic factors through
\begin{equation}
  \frac{\dot H}{H^2} = -(1+q)\,,\qquad
  \frac{\ddot H}{H^3} = j + 3q + 2\,,\qquad
  \frac{\dddot H}{H^4} = s - 4j - 3q(q+4) - 6\,,
\end{equation}
which is the origin of the $q$-, $j$- and $s$-dependent coefficients in
Eqs.~\eqref{eq:g2_obs}--\eqref{eq:g4_obs}.

A useful check on the result is the power-law case $G = G_0 a^{\,n}$, for
which $g_1 = n$ and $g_2 = g_3 = g_4 = 0$ identically, for arbitrary
$q$, $j$ and $s$. Substituting the corresponding
$M_k$ into Eqs.~\eqref{eq:g2_obs}--\eqref{eq:g4_obs} confirms that each
vanishes; omitting the $12\,\dot G^2\ddot G/(G^3H^4)$ term instead leaves a
residual $12n^3(1+q-n)$ in $g_4$.

%=========================================

\end{document}